\documentclass[twocolumn,tighten]{aastex631}
\usepackage[utf8]{inputenc}
\usepackage[T1]{fontenc}
\usepackage{apjfonts}

\usepackage{graphicx}
\usepackage{color, comment}
\usepackage{amsmath,amstext}
\usepackage{ amssymb }
\color{black}
\usepackage{hyperref}
\usepackage{natbib}
\usepackage{hyperref}
\hypersetup{
    colorlinks=true,
    linkcolor=blue,
    filecolor=magenta,      
    urlcolor=cyan,
  }
  \usepackage{footmisc}
  
\begin{document}
\newcommand{\JWST}{\textit{JWST}}
\newcommand{\jwst}{\textit{JWST}}
\newcommand{\HST}{\textit{HST}}
\newcommand{\hst}{\textit{HST}}
\newcommand{\bagpipes}{\textsc{BAGPIPES}}
\newcommand{\eazy}{\textsc{EAzY}}
\newcommand{\sfrten}{\hbox{SFR$_{10}$}}
\newcommand{\sfrcen}{\hbox{SFR$_{100}$}}
\newcommand{\sigmaten}{\hbox{$\sigma_{10}$}}
\newcommand{\sigmacen}{\hbox{$\sigma_{100}$}}

\definecolor{aggiemaroon}{HTML}{500000}
\newcommand{\cjpcomment}[1]{\textcolor{aggiemaroon}{\tt #1}}
\newcommand{\jccomment}[1]{\textcolor{red}{\tt #1}}

\newcommand{\halpha}{\hbox{H$\alpha$}}
\newcommand{\hbeta}{\hbox{H$\beta$}}
\newcommand{\ha}{\hbox{H$\alpha$}}
\newcommand{\hb}{\hbox{H$\beta$}}
\newcommand{\lyalpha}{\hbox{Ly$\alpha$}}
\newcommand{\lya}{\hbox{Ly$\alpha$}}
\newcommand{\neiii}{\hbox{[Ne\,{\sc iii}]}}
\newcommand{\oii}{\hbox{[O\,{\sc ii}]}}
\newcommand{\oiii}{\hbox{[O\,{\sc iii}]}}
\newcommand{\nii}{\hbox{[N\,{\sc ii}]}}
\newcommand{\sii}{\hbox{[S\,{\sc ii}]}}
\newcommand{\siiv}{\hbox{Si\,{\sc iv}}}
\newcommand{\hii}{\hbox{H\,{\sc ii}}}
\newcommand{\heii}{\hbox{He\,{\sc ii}}}

\title{\large  CEERS: Increasing Scatter along the Star-Forming Main Sequence Indicates \\Early Galaxies Form in Bursts}

\correspondingauthor{Justin Cole}
\email{jwc68@tamu.edu}

\author[0000-0002-6348-1900]{Justin W. Cole}
\altaffiliation{NASA FINESST Investigator}
\affiliation{Department of Physics and Astronomy, Texas A\&M
  University, College Station, TX, 77843-4242 USA}
\affiliation{George P.\ and Cynthia Woods Mitchell Institute for
  Fundamental Physics and Astronomy, Texas A\&M University, College
  Station, TX, 77843-4242 USA}
    
\author[0000-0001-7503-8482]{Casey Papovich}
\affiliation{Department of Physics and Astronomy, Texas A\&M University, College
Station, TX, 77843-4242 USA}
\affiliation{George P.\ and Cynthia Woods Mitchell Institute for
 Fundamental Physics and Astronomy, Texas A\&M University, College
 Station, TX, 77843-4242 USA}

\author[0000-0001-8519-1130]{Steven L. Finkelstein}
\affiliation{Department of Astronomy, The University of Texas at Austin, Austin, TX, USA}

\author[0000-0002-9921-9218]{Micaela B. Bagley}
\affiliation{Department of Astronomy, The University of Texas at Austin, Austin, TX, USA}

\author[0000-0001-5414-5131]{Mark Dickinson}
\affiliation{NSF's National Optical-Infrared Astronomy Research Laboratory, 950 N. Cherry Ave., Tucson, AZ 85719, USA}

\author[0000-0001-9298-3523]{Kartheik G. Iyer}
\affiliation{Dunlap Institute for Astronomy \& Astrophysics, University of Toronto, Toronto, ON M5S 3H4, Canada}

\author[0000-0003-3466-035X]{{L. Y. Aaron} {Yung}}
\altaffiliation{NASA Postdoctoral Fellow}
\affiliation{Astrophysics Science Division, NASA Goddard Space Flight Center, 8800 Greenbelt Rd, Greenbelt, MD 20771, USA}
\affiliation{Space Telescope Science Institute, 3700 San Martin Drive, Baltimore, MD 21218, USA}

\author[0000-0003-0541-2891]{Laure Ciesla}
\affiliation{Aix Marseille Univ, CNRS, CNES, LAM, Marseille, France}

\author[0000-0001-5758-1000]{Ricardo O. Amor\'{i}n}
\affiliation{ARAID Foundation. Centro de Estudios de F\'{\i}sica del Cosmos de Arag\'{o}n (CEFCA), Unidad Asociada al CSIC, Plaza San Juan 1, E--44001 Teruel, Spain}
\affiliation{Departamento de Astronom\'{i}a, Universidad de La Serena, Av. Juan Cisternas 1200 Norte, La Serena 1720236, Chile}

\author[0000-0002-7959-8783]{Pablo Arrabal Haro}
\affiliation{NSF's National Optical-Infrared Astronomy Research Laboratory, 950 N. Cherry Ave., Tucson, AZ 85719, USA}

\author[0000-0003-0883-2226]{Rachana Bhatawdekar}
\affiliation{European Space Agency (ESA), European Space Astronomy Centre (ESAC), Camino Bajo del Castillo s/n, 28692 Villanueva de la Cañada, Madrid, Spain}

\author[0000-0003-2536-1614]{Antonello Calabr\`o}
\affiliation{INAF Osservatorio Astronomico di Roma, Via Frascati 33, 00078 Monteporzio Catone, Rome, Italy}

\author[0000-0001-7151-009X]{Nikko J. Cleri}
\affiliation{Department of Physics and Astronomy, Texas A\&M University, College Station, TX, 77843-4242 USA}
\affiliation{George P.\ and Cynthia Woods Mitchell Institute for Fundamental Physics and Astronomy, Texas A\&M University, College Station, TX, 77843-4242 USA}

\author[0000-0002-6219-5558]{Alexander de la Vega}
\affiliation{Department of Physics and Astronomy, University of California, 900 University Ave, Riverside, CA 92521, USA}

\author[0000-0003-4174-0374]{Avishai Dekel}
\affil{Racah Institute of Physics, The Hebrew University of Jerusalem, Jerusalem 91904, Israel}

\author[0000-0003-4564-2771]{Ryan Endsley}
\affiliation{Department of Astronomy, The University of Texas at Austin, Austin, TX, USA}

\author[0000-0003-1530-8713]{Eric Gawiser}
\affiliation{Department of Physics and Astronomy, Rutgers, the State University of New Jersey, Piscataway, NJ 08854, USA}

\author[0000-0002-7831-8751]{Mauro Giavalisco}
\affiliation{University of Massachusetts Amherst, 710 North Pleasant Street, Amherst, MA 01003-9305, USA}

\author[0000-0001-6145-5090]{Nimish P. Hathi}
\affiliation{Space Telescope Science Institute, 3700 San Martin Drive, Baltimore, MD 21218, USA}

\author[0000-0002-3301-3321]{Michaela Hirschmann}
\affiliation{Institute of Physics, Laboratory for galaxy evolution, EPFL, Observatory of Sauverny, Chemin Pegasi 51, 1290 Versoix, Switzerland}
\affiliation{INAF, Astronomical Observatory of Trieste, Via G.P. Tiepolo 11, 34134 Trieste, Italy}

\author[0000-0002-4884-6756]{Benne W. Holwerda}
\affiliation{Physics \& Astronomy Department, University of Louisville, 40292 KY, Louisville, USA}

\author[0000-0001-9187-3605]{Jeyhan S. Kartaltepe}
\affiliation{Laboratory for Multiwavelength Astrophysics, School of Physics and Astronomy, Rochester Institute of Technology, 84 Lomb Memorial Drive, Rochester, NY 14623, USA}

\author[0000-0002-6610-2048]{Anton M. Koekemoer}
\affiliation{Space Telescope Science Institute, 3700 San Martin Drive, Baltimore, MD 21218, USA}

\author[0000-0003-1581-7825]{Ray A. Lucas}
\affiliation{Space Telescope Science Institute, 3700 San Martin Drive, Baltimore, MD 21218, USA}
    
\author[0000-0002-9572-7813]{Sara Mascia}
\affiliation{INAF Osservatorio Astronomico di Roma, Via Frascati 33, 00078 Monteporzio Catone, Rome, Italy}

\author[0000-0001-5846-4404]{Bahram Mobasher}
\affiliation{Department of Physics and Astronomy, University of California, 900 University Ave, Riverside, CA 92521, USA}

\author[0000-0003-4528-5639]{Pablo G. P\'erez-Gonz\'alez}
\affiliation{Centro de Astrobiolog\'{\i}a (CAB), CSIC-INTA, Ctra. de Ajalvir km 4, Torrej\'on de Ardoz, E-28850, Madrid, Spain}

\author[0000-0002-9415-2296]{Giulia Rodighiero}
\affiliation{Department of Physics and Astronomy, Università degli Studi di Padova, Vicolo dell’Osservatorio 3, I-35122, Padova, Italy}
\affiliation{INAF - Osservatorio Astronomico di Padova, Vicolo dell’Osservatorio 5, I-35122, Padova, Italy}

\author[0000-0001-5749-5452]{Kaila Ronayne}
\affiliation{Department of Physics and Astronomy, Texas A\&M University, College Station, TX, 77843-4242 USA}
\affiliation{George P.\ and Cynthia Woods Mitchell Institute for Fundamental Physics and Astronomy, Texas A\&M University, College Station, TX, 77843-4242 USA}

\author[0000-0002-8224-4505]{Sandro Tacchella} \affiliation{Kavli Institute for Cosmology, University of Cambridge, Madingley Road, Cambridge CB3 0HA, UK} \affiliation{Cavendish Laboratory, University of Cambridge, 19 JJ Thomson Avenue, Cambridge CB3 0HE, UK}

\author[0000-0001-6065-7483]{Benjamin J. Weiner}
\affiliation{MMT/Steward Observatory, University of Arizona, 933 N. Cherry St, Tucson, AZ 85721, USA}

\author[0000-0003-3903-6935]{Stephen M.~Wilkins} %
\affiliation{Astronomy Centre, University of Sussex, Falmer, Brighton BN1 9QH, UK}
\affiliation{Institute of Space Sciences and Astronomy, University of Malta, Msida MSD 2080, Malta}

\begin{abstract}
%
We present the star-formation-rate---stellar-mass (SFR-M$_\ast$) relation for galaxies in the CEERS survey at $4.5\leq z\leq 12$.  We model the \jwst\ and \hst\ rest-UV and rest-optical photometry of galaxies with flexible star-formation histories (SFHs) using \bagpipes.  We consider SFRs averaged from the SFHs over 10~Myr (\sfrten) and 100~Myr (\sfrcen), where the photometry probes SFRs on these timescales, effectively tracing nebular emission lines in the rest-optical (on $\sim10$~Myr timescales) and the UV/optical continuum (on $\sim100$ Myr timescales). We measure the slope, normalization and intrinsic scatter of the SFR--M$_\ast$ relation, taking into account the uncertainty and the covariance of galaxy SFRs and $M_\ast$.    From $z\sim 5-9$ there is larger scatter in the $\sfrten-M_\ast$ relation, with $\sigma(\log \sfrcen)=0.4$~dex,  compared to the $\sfrcen-M_\ast$ relation, with $\sigma(\log \sfrten)=0.1$~dex.    This scatter increases with redshift and increasing stellar mass, at least out to $z\sim 7$.  These results can be explained if galaxies at higher redshift experience an increase in star-formation variability and form primarily in short, active periods, followed by a lull in star formation (i.e. ``napping'' phases). We see a significant trend in the ratio $R_\mathrm{SFR}=\sfrten/\sfrcen$ in which, on average, $R_\mathrm{SFR}$ decreases with increasing stellar mass and increasing redshift. This yields a star--formation ``duty cycle'' of $\sim40\%$ for galaxies with $\log M_\ast/M_\odot\geq 9.3$, at $z\sim5$, declining to $\sim20\%$ at $z\sim9$. Galaxies also experience longer lulls in star formation at higher redshift and at higher stellar mass, such that galaxies transition from periods of higher SFR variability at $z\gtrsim~6$ to smoother SFR evolution at $z\lesssim~4.5$.
%
%
\end{abstract}
%
%
\keywords{High-redshift galaxies (734), Emission line galaxies (459), Galaxy evolution (594), Post-starburst galaxies (2176), Starburst galaxies (1570), Star formation (1569)}
\section{Introduction}
The beginning of galaxy formation and the early build-up of stellar mass in galaxies remains an unsolved area of astrophysical research. These are relevant both to understand the conditions that led to the onset of star formation in galaxies, and it is important to understand when and how these galaxies reionized the intergalactic medium (IGM). Understanding early galaxy formation and stellar mass build-up has become even more pertinent in the era of \JWST, where its data have identified via photometry, and confirmed via spectroscopy, galaxies at $z>7$ and as high as $z\sim12$ \citep[see for example, ][]{finkelstein:22b, arrabal_haro:23, curtis-lake:23, labbe:23a}.  In particular, \jwst\ has moved the study of these early galaxies to the point where we can characterize their properties from large, statistical samples with exquisite data.

From redshifts $z\sim1-6$, the star-formation history of galaxies appears relatively smooth.
The cosmic star-formation-rate (SFR) density grows from the earliest redshifts ($z \gtrsim 6$), peaks around $z\sim2$, and is followed by a slow decline to the present, $z\sim 0$ \citep{madau_dickinson:14}.   This evolution is further evidenced by the evolution of the relation between between the SFR and stellar mass of galaxies \citep[the so-called ``star forming--main-sequence,'' or SFMS, ][]{noeske:07, whitaker:12, speagle:14, salmon:15, iyer:18, curtis-lake:21, popesso:23}. Similarly, the interpretation of observational data has argued that star formation histories (SFHs) of individual galaxies follow this trend, albeit  with a dependence on mass, such that more massive galaxies start and peak at earlier cosmic times \citep{papovich:11, papovich:15, pforr:12, reddy:12, speagle:14, pacifici:16a,iyer:19}. Therefore, from ``cosmic noon'' (the peak of the cosmic SFR density at $z\sim2$) to the highest redshifts we have explored, galaxy formation has appeared relatively smooth and orderly for galaxies, at least as a population.  

However, simulations of galaxy formation find that while on average the SFHs of galaxies are smooth \citep[e.g.,][]{finlator:11,wilkins:17,wilkins:23a,bird:22}, they are much more variable on shorter timescales, with larger variations in the SFR \citep{finlator:06,dave:16,ma_x:16,katz:23b,sun:23}.   \citet{sun:23} recently illustrated this with simulations showing that galaxies at higher redshifts experience greater variability in their specific SFRs (sSFR $\equiv$ SFR/$M_\ast$).  In these simulations, this stems from the fact that galaxies are more compact, with higher gas-- and stellar-mass surface densities.  This leads to higher feedback (per unit mass) from supernovae and winds from massive stars, which more readily expel gas, and disrupt gas accretion.  These feedback mechanisms disrupt gas accretion, and subsequently create a lull in the star formation.  This naturally predicts more galaxies in ``burst'' modes of star formation (with elevated sSFR) and galaxies with temporary cessation of star-formation \citep[so called ``napping,'' or ``mini-quenched,'' galaxies, ][]{looser:23a, looser:23b, dome:24}. The timescales of this feedback relative to the star-formation, and its amplitude (i.e., strength per unit mass) have a strong impact on the SFR and sSFR \citep[e.g.,][]{furlanetto:22}.   

With \jwst, there has been renewed interest in the SFR variability in high-redshift galaxies.  
\citet{endsley:23a, whitler:23} argued that higher variability in the SFRs of galaxies is needed to explain the increase in the frequency of galaxies with evidence for strong emission-line equivalent-widths (EW).  Other recent work from \jwst\ imaging of low-mass  ($10^8$~$M_\odot$) galaxies at $z\sim 5-6$ shows that $\sim$60\% of them have \ha/UV-flux ratios that indicate they are experience bursts and variability in their SFHs \citep{asada:23}.  Yet other studies have invoked a higher variability in the SFRs of high redshift galaxies to explain the higher-than-expected number density of galaxies measured at the bright end of the rest-UV luminosity functions \citep[i.e.,][]{finkelstein:23, mason:23}.  This is also seen in updated simulations of galaxy formation \citep[e.g.,][]{shen:23,wilkins:23a}, and inferred from studies of the evolution of galaxies along the SFR--$M_\ast$ relation \citep[the SFR ``gradient'', e.g.,][]{ciesla:23}.   

Motivated by these observations and apparent problems, here we study the evolution of the SFR--$M_\ast$ plane using \jwst\ data from the Cosmic Evolution Early Release Science (CEERS) survey. CEERS is a \jwst\ survey of the Extended Groth Strip \citep[EGS, ][]{ilbert:13}, providing seven photometric bands with NIRCam covering $\sim1 - 5~\mu$m.  This field also has deep legacy imaging from \HST/ACS and WFC3 ($\sim0.6-1.6~\mu$m) and other ground-based and space telescopes \citep[see,][]{grogin:11,koekemoer:11,stefanon:17}. We model the \jwst/NIRCam and \hst\ multi-wavelength photometry with stellar population models using flexible SFHs to derive constraints on the stellar masses, SFRs and redshifts.  We then use the full posteriors from these fits to study the evolution of the slope, normalization and intrinsic scatter of the SFR--$M_\ast$ relation (the ``star forming--main-sequence''), and the scatter of the SFR as a function of galaxy mass and cosmic time (redshift).  We invoke a novel method to measure the evolution of the SFMS that accounts for the uncertainties and covariance from the posteriors of all the variables. 

The outline for the rest of this paper is as follows. Section \ref{section:data} presents the CEERS data used for this analysis and discusses our selection methods. Section \ref{section:SED} provides details of our analysis methods used to model the galaxy spectral energy distributions (SEDs) with stellar population models and flexible SFHs.  We also discuss the ability of the modeling to distinguish between the SFR on timescales of 10 and 100~Myr. In Section \ref{section:results}, we present measurements of the SFMS and its evolution.  We provide a discussion of the implications of our results and compare the results to theoretical predictions in Section \ref{section:discussion}.  In section \ref{section:etfin} we present our conclusions.  We also include an Appendix (Section \ref{section:appendix}) with tests of our ability to fit and recover the scatter of the SFMS using our methods.  Throughout this paper, we assume a flat-$\Lambda$CDM cosmology with values from Planck ($\mathrm{H}_0 = 67.36$, $\Omega_\mathrm{M} = 0.3158$). All magnitudes  are presented in the AB system \citep{Oke1983, fukugita1996}. We assume a \citet{chabrier:03} initial mass function (IMF) for derived quantities such as the SFR and stellar mass. 

\section{Observations and Data Selection}\label{section:data}

We leverage the CEERS dataset \citep{finkelstein:22b,bagley:23}, providing \JWST\ observations in the CANDELS \citep{koekemoer:11, grogin:11} EGS field.   The dataset includes NIRCam imaging from \jwst, and ACS and WFC3 imaging from \hst.    The detailed description of the NIRCam data reduction and methods for extracting the photometry can be found in \citet{bagley:23}.

Our work utilizes the ten NIRCam pointings obtained by CEERS, with imaging in the F115W, F150W, F200W, F277W, F356W, F410M, and F444W filters, spanning a wavelength range from approximately 1 to 5 microns.  The \HST\ data includes imaging from the ACS F606W and F814W bands, and the WFC3 F125W, F140W, and F160W bands (with some data in F105W also available), covering 0.6--1.6 micron.   The NIRCam data are aligned astrometrically to the \HST\ imaging \citep[see,][]{finkelstein:23}.

We adopt the photometric catalog from \citet{finkelstein:23b}.   Galaxies are detected using the F277W imaging, and photometry is measured using matched apertures applied to the NIRCam images and \HST\ ACS and WFC3 images.  The catalog then includes photometric data points from the 12 bandpasses spanning a wavelength range of 0.6 to 5 $\mu$m for $\approx 100,000$ sources.  In addition, we also make use of \hst/WFC3 F105W-band imaging that covers approximately 20\% of the CEERS/NIRCam field.  This provides an additional photometric point at $\sim$1~$\mu$m for $\sim20\%$ of the galaxies in the catalog.  Below we discuss the selection criteria applied to construct the sample used in this study (Sections \ref{section:sample} and \ref{section:removal}) with additional explanations for sample selection provided in Section \ref{section:SED}.

\subsection{The CEERS $4.5 \leq z \leq 12$ Sample}
\label{section:sample}

We selected galaxies with $m_{F277W} \leq 28.2$, which is one magnitude brighter than the 5-$\sigma$, point-source detection limit \citep[$\approx 29.2$ mag, ][]{finkelstein:23, bagley:23}.  This ensures our sample has high fidelity through the benefit of higher SNR in multiple photometric bands. We also require galaxies to have photometric redshifts between $z_{peak}\sim4.5$ and $z_{peak}\sim12$, using redshifts ($z_{peak}$) derived from \eazy\ \citep{brammer:08}. A description of the \eazy-derived redshifts can be found in \citet{finkelstein:23b}.  The lower redshift bound, $z > 4.5$, ensures that the photometric catalog covers the wavelength region around the the Lyman-limit/\lya-breaks for all galaxies, allowing for more accurate photometric redshifts. At the upper redshift bound, $z~\leq~12$, the photometric catalog covers wavelengths around the Balmer--limit break (3646~\AA) and the 4000~\AA--break (from line blanketing of metals in stellar photospheres).  Therefore, our selection of $4.5 < z_{peak} \leq 12$ provides rest-frame far-UV to rest-optical photometry for the full sample, and this enables higher fidelity in derived quantities, such as the stellar mass and SFR estimates from SED fitting \citep{pforr:12}. Applying the magnitude limit and redshift-range selection to the catalog yields a parent sample of 5,669 sources.  We then inspected this sample visually and identified several sources of potential contamination, which we screened through the application of additional selection criteria, described below. 
\subsection{Matching HST Photometry, and Outlier Removal}
\label{section:removal}
We use three additional selection criteria to flag and exclude sources that stem from photometric contamination effects, and we also remove sources that have contamination from possible AGN.   First, because galaxies are matched to the legacy \HST\ imaging, we require that the photometric measurements between the \hst\ data and \jwst\ data be consistent. Specifically we require, 
\begin{equation}
\label{eq:mag}
    \frac{|m(\mathrm{F150W}) - m(\mathrm{F160W})|}{\sqrt{\sigma_\mathrm{F150W}^2 + \sigma_\mathrm{F160W}^2}} \leq 2.5, 
\end{equation}
where $m(\mathrm{F150W})$ ($~m(\mathrm{F160W})~$) and $\sigma_\mathrm{F150W}$ ($\sigma_\mathrm{F160W}$) are the measured photometry and uncertainty in the \jwst/NIRCam F150W (\hst/WFC3 F160W) bandpass.  This requirement forces the \HST/F160W photometry to be within 2.5-$\sigma$ of the \JWST/F150W photometry.  In practice, we found that this selection requirement flags and removes obvious contamination between foreground sources, often because of differences in the point spread function (PSF) between \HST\ and \JWST\ (e.g., objects with optical diffraction spikes). This also flagged and removed several spurious imaging artifacts and objects near the edges of the images.   This selection removed 1231 galaxies (21\% of 5669) from the initial parent sample. 

Second, to ensure a robust, high-quality sample of galaxies, we added a requirement that the galaxies in our sample have well-defined photometric redshifts.  Specifically, we exclude objects with multiple ``peaks'' in the photometric redshift probability density functions (PDFs, $P(z)$) or that have a very broad $P(z)$.  In such situations, the uncertainty in the redshift $P(z)$ translates to very large uncertainties in the derived physical parameters like masses and SFRs \citep[see, for example ][]{mobasher:07, darvish:15, mobasher:15}.  As the goal of our study is to understand the latter, we exclude objects where the photomoetric-redshift accuracy would be the limiting factor.   To make this selection, we defined a redshift, $P(z)$ ``\textsc{width\_flag}" using the \eazy\ redshift $P(z)$ functions for each source in the catalog. We start by using a method similar to the highest-density interval \citep[e.g.,][]{bailer-jones:18}.  We start at the peak probability of the PDF and move away from the peak by stepping through redshift space until the area under the $P(z)$ curve integrates to 0.7 (roughly analogous to the inner 16th--84th percentile).  This then defines a $z_\mathrm{low}$ and $z_\mathrm{high}$ that  encompass 70\% of the probability distribution. We require that the difference between $z_\mathrm{low}$ and $z_\mathrm{high}$ meets the criteria
\begin{equation} \label{eq:selection}
    \frac{z_{high} - z_{low}}{(1+z_{peak})} \leq 0.5.
\end{equation}
This ensures that we are only using galaxies that have relatively well-defined redshift probability spaces because 70\% of the integrated $P(z)$ lies within $0.5 \times (1+z_\mathrm{peak})$. 

This step removes 46\% of the initial sample; however, it leaves galaxies with much more robust photometric redshifts, which translates to a higher-quality sample of objects.   Nevertheless, we have also verified that our results and conclusions change negligibly if we \textit{include} the objects rejected by Equation~\ref{eq:selection}, at a $\lesssim 15\%$ level for median values and even smaller affects on the inferred parameter errors.  We interpret this as implying the objects with poor $P(z)$ constraints also have poor constraints on quantities such as stellar mass and SFR \citep{mobasher:07, darvish:15, mobasher:15}.  Objects with poorly defined likelihoods matter less when we use the fully marginalized likelihoods (i.e., $P(M_\ast)$, $P(\mathrm{SFR})$) and the joint likelihood (i.e.,g $P(M_\ast, \mathrm{SFR})$) to model the SFMS relation and its scatter (as we do in our analysis below).   For this reason we present results using the sample that excludes objects that do not satisfy Equation~\ref{eq:selection}, as they are the most robust. 

Lastly, we also reject galaxies that may contain an AGN.   We used a \textit{Chandra}-based X-ray source catalog \citep{nandra:15} of the EGS field to determine if any galaxies in our sample have high X-ray luminosities \citep[$L_X / \mathrm{erg s}^{-1} \geq 3\times10^{42}$, ][]{xueyq:16,luo:16}.  We matched these sources to our catalog using a search radius of 1\arcsec, which yield only 3 sources with $L_X \geq 3\times10^{42}$ erg~s$^{-1}$, which we excluded from the sample.    Although this does not account for dust-obscured AGN, or faint AGN, by rejecting X-ray--bright AGN, this improves the likelihood that the galaxies in our sample have rest-UV--optical SEDs that are dominated by their stellar light and/or nebular emission from star-forming regions.

Following all the selection criteria, our final sample includes 1863 galaxies.  We then proceed to model the galaxies' SEDs using stellar populations fits with the \bagpipes\ code. 

\section{Spectral Energy Distribution Fitting} 
\label{section:SED}
\subsection{\bagpipes: priors and parameter spaces}

We fit the SEDs of all sources in our sample using \bagpipes\ \citep[v1.0.1, ][]{carnall:18}, a Bayesian SED-fitting code designed to fit photometry and/or spectra of galaxies with a broad range of parameter spaces. \bagpipes\ utilizes a chi-squared--likelihood function, assuming that photometric errors follow a normal distribution, and allows for a wide range in the prior parameter spaces. \bagpipes\ then samples the posteriors for model parameters using the \textsc{MultiNest} nested sampling algorithm (see \citealt{feroz:09,carnall:18}). Table~\ref{table:SED_fitting} provides a summary of the parameters, their range, prior, and other information.   We provide more details in here.  

With \bagpipes, we fit all galaxies assuming a \citet{chabrier:03} IMF and we allow \bagpipes\ to fit for $\log(M_*/M_\odot) \in [6,15]$ with a uniform prior, providing a wide prior space for stellar mass measurements. We also assume a \citet{calzetti:00} dust law, which has been suggested to be representative of high-redshift galaxies \citep{bowler:23}. We employ Binary Population and Spectral Synthesis \citep[BPASS v2.2.1, ][]{eldridge:17} stellar templates, and fit over a range of metallicity (Z~/~Z$_\odot$) $\in [0.001,1]$.   We adopt the BPASS models that include the effects of mass-transfer between binary stars with an upper limit of 300~$M_\odot$ to better reproduce the emission-line strength and ratios seen in some galaxies \citep[e.g.,][]{steidel:18,strom:17,larson:22,olivier:22}.  
We further allow for nebular emission \citep[modeled with \textsc{Cloudy},][]{ferland:17} where the metallicity of the nebular gas is equal to that of the stellar population ($Z_{\mathrm{stars}} = Z_\mathrm{neb}$). The ionization parameter ranges over $\log{(U)} \in [-4,-1]$ and we use a log-uniform prior.  

\begin{deluxetable}{cc}[t!]
\caption{Parameters, parameter spaces and priors used in our custom version of \bagpipes. }
\tablehead{\colhead{Parameter} & \colhead{Parameter Space (Prior)\tablenotemark{\dag}}
\label{table:SED_fitting}}
\startdata
    SFH: Gaussian Process & 4 time bins: $t_X$ \\
    DenseBasis & (Dirichlet continuity) \\
    \citep{iyer:19} & $\log{(SFR)} \in [-3,4]$ \\
    & (Uniform in $\log_{10}$) \\
    \hline
    Stellar Mass: & $\log{(M_*/M_\odot)} \in [6,15]$ \\
    & (Uniform) \\
    \hline
    Redshift: & $z \in [0,20]$\\
    & (\eazy\ PDF) \\
    \hline
    Ionization Parameter: & $\log{(U)} \in [-4,-1]$\\
    & (Uniform) \\
    \hline
    Metallicity: & $Z~/~Z_\odot \in [0.001,1]$, $Z_\mathrm{stellar}=Z_\mathrm{neb}$\\
    & (Uniform) \\
    \hline
    \citet{calzetti:00} Dust Law: & $A_V / \mathrm{mag}\in [0,6]$\\
    & (Uniform) \\
    \hline
    \enddata
    \tablenotetext{\dag}{Parameter spaces are shown in brackets and the prior imposed on each parameter space is shown in parentheses below each parameter space.}
\end{deluxetable}

\begin{figure*}[t!]
    \centering
    \includegraphics[width=\linewidth]{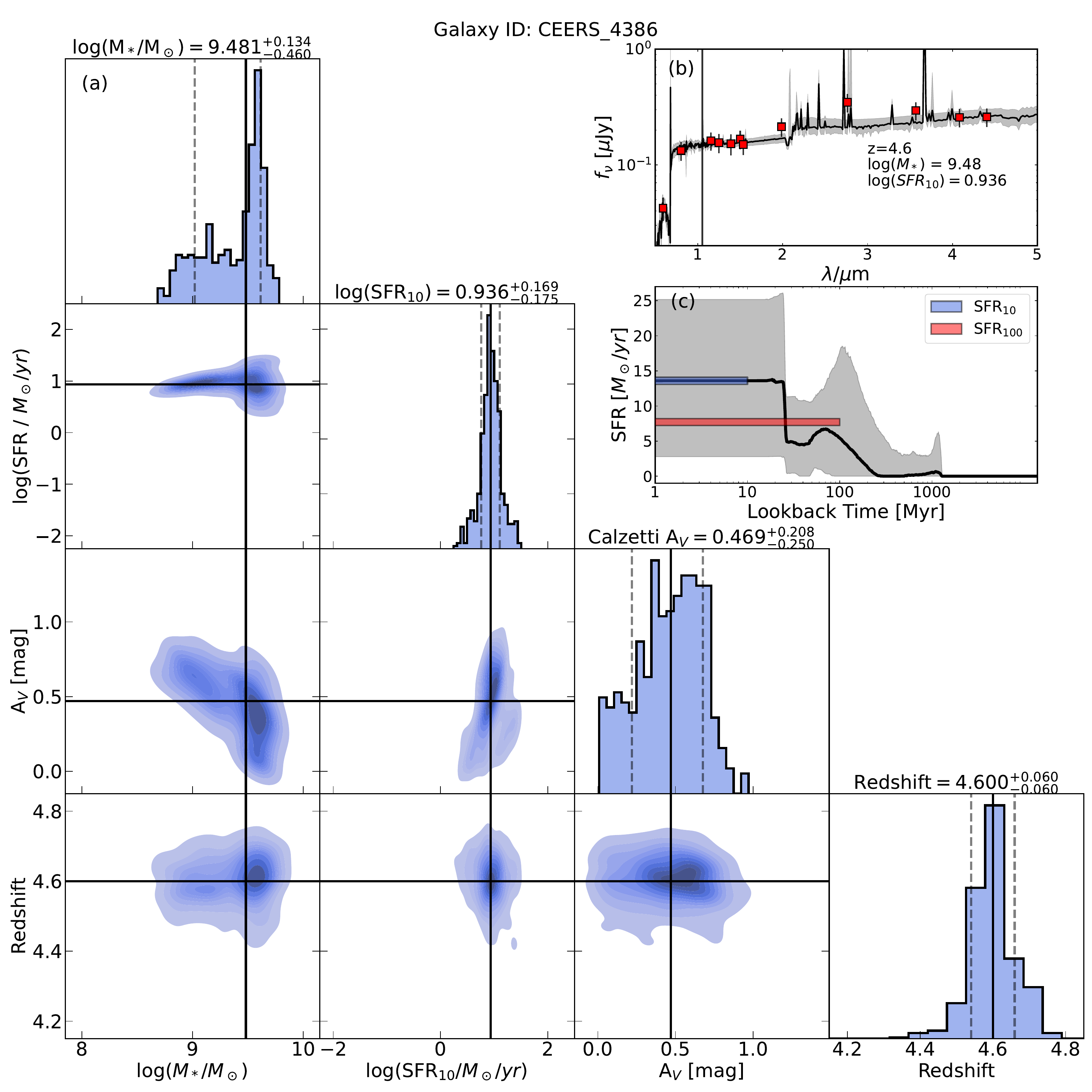}
    \caption{An example of the posterior space and best-fit models for a representative galaxy in our sample (CEERS\_4386). The panels in (a) are the corner plot, which shows the 2-D kernel density estimates of the multi-dimensional posterior space for stellar mass, the star formation rate (averaged over the previous 10 ~Myr, \sfrten), dust attenuation ($A_V$), and redshift estimated from the model fits. Panel (b) shows the median, best-fit SED as a solid black line with the inner-$68^{th}$ percentile as the gray region.  The red squares and error bars show the measured photometry and uncertainties. The median  redshift, stellar mass and \sfrten\ for this galaxy are listed in the panel inset. Panel (c) shows the inferred star formation history (SFH), with the median SFH as a solid black line and the inner-$68^{th}$ percentile as the gray shaded region.  The blue and red rectangles show the SFR averaged over the previous 10 and 100~Myr, respectively (\sfrten\ and \sfrcen, respectively).  }
    \label{fig:soloSED}
\end{figure*}

For the SFH parameterization in \bagpipes, we adopt the Gaussian Process (GP) model from \textsc{densebasis} \citep{iyer:19}. For a more detailed description of the GP implementation, see \citet{carnall:19a}.
In this case, the SFH is defined by $N$ parameters, \{$t_X$\}, where $X=1\ldots N$.  Each \{$t_X$\} is the lookback time at which a fraction, $1/N$, of the total stellar mass is formed.  The value for each $t_X$ is allowed to vary in the fit to force this condition. \citet{iyer:19} has demonstrated the robustness of this methodology in reconstructed bursty SFHs, which may be more indicative of star formation in the early Universe \citep{kimm:15, ceverino:18, barrow:20, furlanetto:22, kannan:22, endsley:23a, sun:23}. 

For this work, we use $N=4$ and fit for \{$t_X$\} such that each lookback time bin contains $1/4$ = 25\% of the total stellar mass formed and we set a Dirichlet Continuity prior \citep[see][for a more detailed description]{leja:17,leja:19a,iyer:19} on the SFH, forcing continuity across the time bins.  Regarding the choice of $N=4$, as we discuss in Section \ref{section:sfrs}, the photometric data are sensitive to star formation on roughly three different timescales, which sets $N\geq 3$:  (1) the data measuring the strength of nebular emission (e.g., \oiii, \hb, \ha) probe $<$10~Myr timescales (the lifetime of O-type stars); (2) the data measuring the rest-UV probe  $\sim$100~Myr timescales (e.g., \citealt{kennicutt:98}); (3) the data measuring the region of the 4000~\AA/Balmer-break probe timescales of $>$500~Myr (the lifetimes of A-type and later-type stars).  We set $N=4$ to include these three time bins, and to allow for earlier episodes of star-formation lost in the glare of the current stars \citep[e.g., the ``outshining'' effect,][]{conroy:13,papovich:23}.    We have tested our methodology using $N=3$ and $N=6$ with negligible differences between inferred parameters, and therefore this choice does not impact our results.    For the individual time bins, we employ a log-uniform prior for $\log{\left(\mathrm{SFR}/M_\odot/yr^{-1}\right)} \in [-3,4]$, which allows for a wide range of SFRs for each time bin.

Modifying the method presented in \citet{chworowsky:23b}, we have adapted \bagpipes\ to use the  $P(z)$ distributions as a prior on the redshift for each galaxy (this method was also implemented by \citealt{papovich:23}).  As mentioned in Section \ref{section:removal}, we use the $P(z)$ for each source measured with \eazy. We opt to use this methodology as it provides a more physically-motivated, and therefore more meaningful, distribution from which to draw redshift probabilities. 

We performed the SED fitting using \bagpipes\ on the Texas A\&M High Performance Research Computing (HPRC) Grace cluster\footnote{See \url{https://hprc.tamu.edu/kb/User-Guides/Grace}, where the namesake of the cluster is  \href{https://en.wikipedia.org/wiki/Grace_Hopper}{Grace Hopper}.}, which has 800 Intel 6248R 3.0GHz 24-core processors with (at least) 384 Gb RAM. On a single core, the clock time to fit each galaxy ranged from 4 to 10 minutes.  The total clock time to fit the full sample was $\sim 520$ core--hours (though the galaxies were fit in parallel so the wall-clock time was substantially shorter).  

\begin{figure*}[t!]
    \centering
    \includegraphics[width=0.9\linewidth]{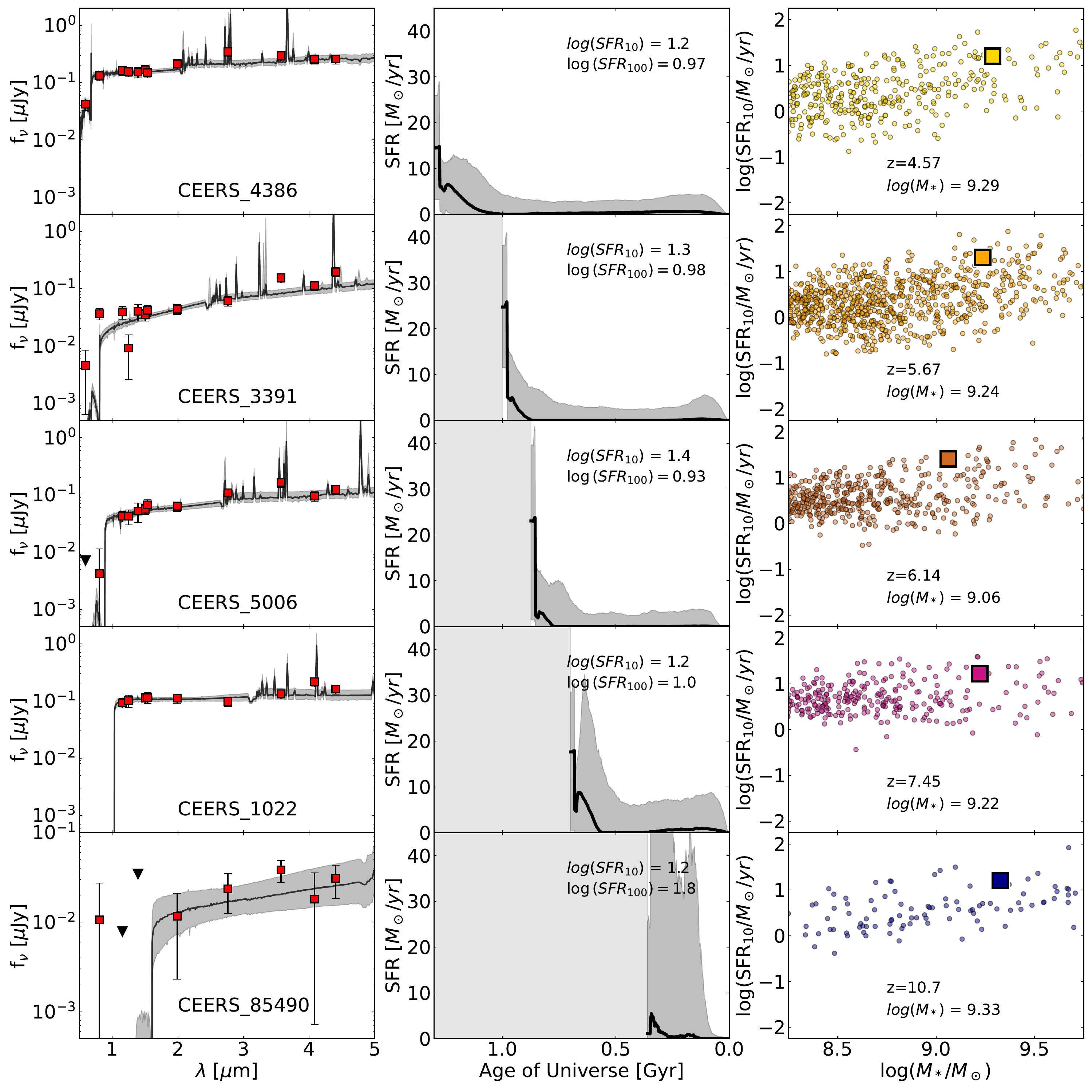}
    \caption{Example SEDs for galaxies with log($M_*/M_\odot$) $\sim 9$ and log(\sfrten) $\gtrsim 1.0$, such that they lie above the median SFR at their stellar mass compared to the SFMS.   Each row shows one galaxy selected from each of the five redshift bins (increasing redshift from top to bottom). In the left panel of each row, the red squares and error bars show the observed photometry and uncertainties. Photometric points with SNR $<$1 are shown as black, downward-facing triangles to indicate upper limits.  The median best-fit model SED for each galaxy is shown as the black solid line and the inner-68$^{th}$ percentile range as the gray-shaded region.  The middle panel in each row shows the model SFH with the median (black line) and inter-68 percentile range for each galaxy.  In the right panel of each row,  we show the location of each galaxy (large square) in the SFR--$M_\ast$ plane compared to the other galaxies (small symbols) in the same redshift bin. In all cases (especially for $ z < 9$), the galaxy SEDs show indications of strong emission lines which causes an increase in the SFH in the previous $\sim$10 Myr compared to the past average. }
    \label{fig:highsfr}
  \end{figure*}

\begin{figure*}[t!]
    \centering
    \includegraphics[width=0.9\linewidth]{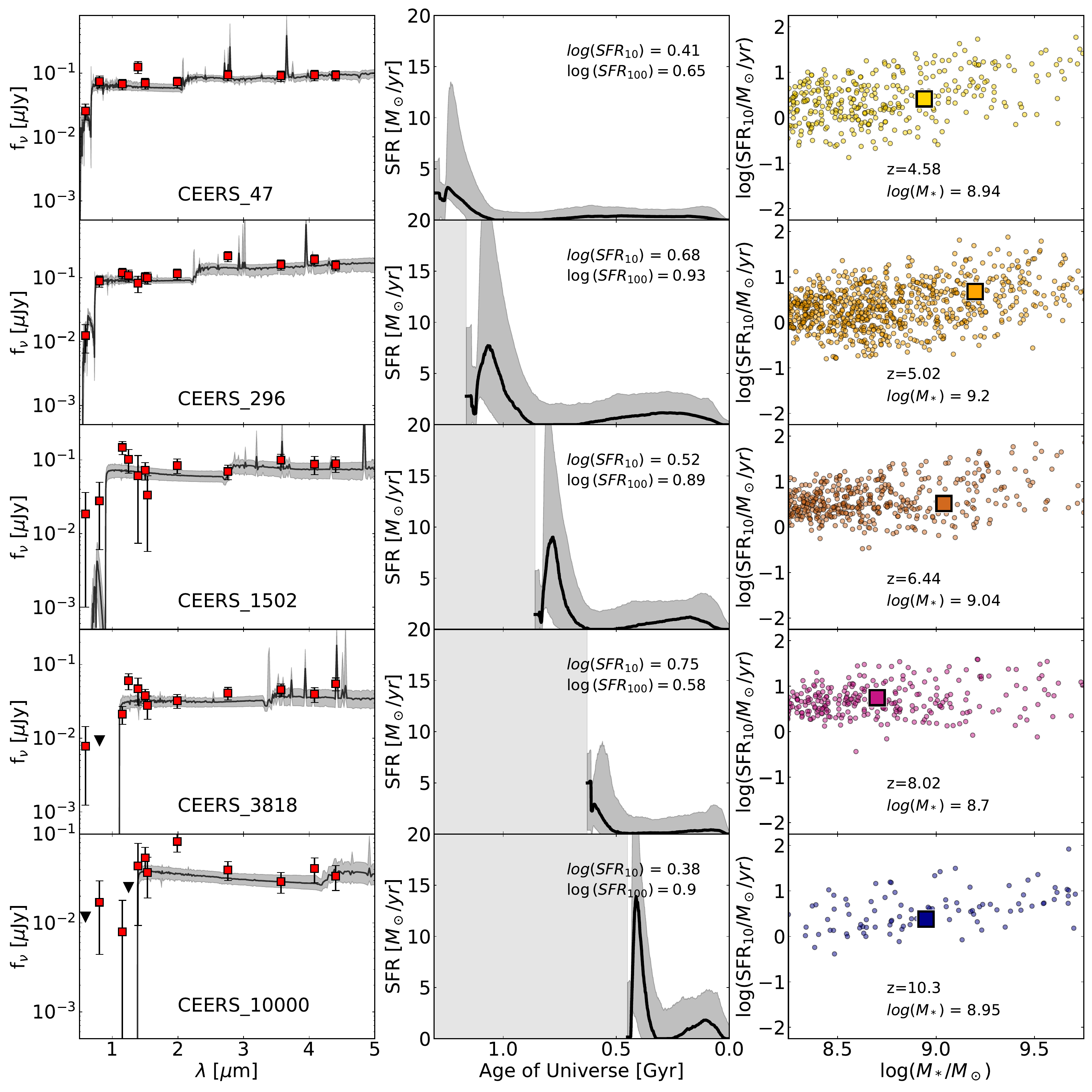}
    \caption{The same as in Figure \ref{fig:highsfr}, but here we show example galaxies that have SFRs within 0.1 dex of the median SFR at their redshift.  In these cases, the SFR over the past 10~Myr (i.e., \sfrten) is consistent with the past average over the previous 100~Myr (\sfrcen).}
    \label{fig:sfms_SED}
\end{figure*}

\begin{figure*}[t!]
    \centering
    \includegraphics[width=0.9\linewidth]{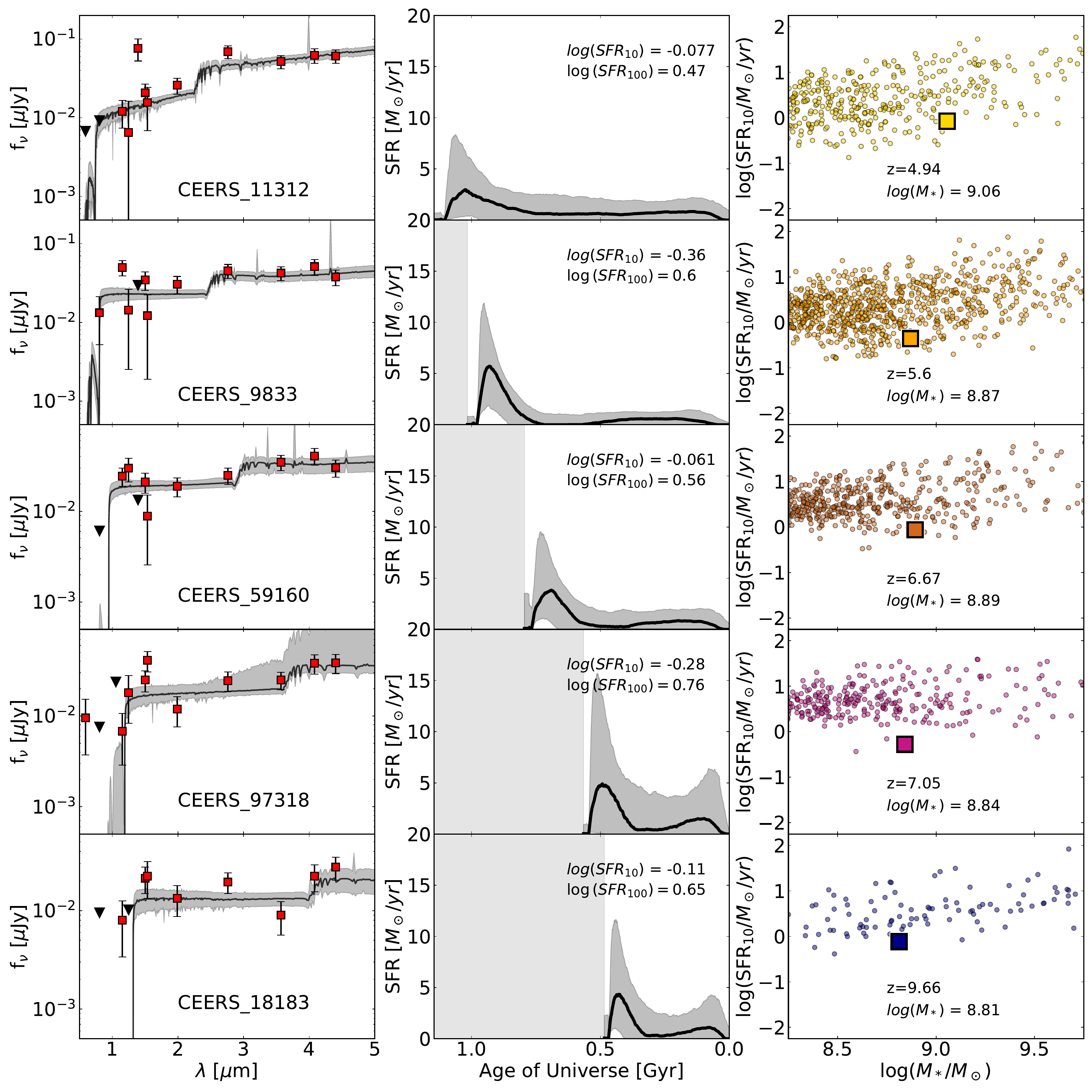}
    \caption{The same as in Figure \ref{fig:highsfr}, but for galaxies with log(\sfrten/$M_\odot$~yr$^{-1}$) $\lesssim 0$, below the median SFR of the SFR--$M_*$ plane.  In these cases, the lack of evidence for strong emission lines in the SEDs implies the SFR has declined over the previous $\sim$10~Myr compared to the past average. }
    \label{fig:lowsfr}
\end{figure*}
  
Figure \ref{fig:soloSED} shows the  results from the fits to one of our galaxies, CEERS\_4386, which has a stellar mass representative of the mean stellar mass of the sample and high SFR compared to galaxies at $z\sim5$.
The corner plot (panel (a))  shows the posterior space for the stellar mass (accounting for mass-loss due to stellar evolution), SFR, dust attenuation ($A_V$) and redshift using two-dimensional kernel density estimates (KDEs, where darker colored layers show more densely populated regions of the posterior spaces).   The inset panels of Figure \ref{fig:soloSED} show the fitted model SED (panel (b)) and the constraints on the SFH (panel (c)).  Importantly, the thick blue and red lines in the latter show the SFR for the galaxy derived by averaging the SFH over the previous 10 Myr and 100 Myr timescales, respectively.  This will become important as the SED fitting is sensitive to features that vary on these timescales:   the strength of the nebular emission responds to changes in the SFR over timescales of $\sim$5 Myr (the lifetime of a low--metallicity, 25~$M_\odot$ O star), while the strength of the rest-frame UV continuum responds to changes in the SFR over timescales of $\sim$20-100 Myr \citep[the lifetime of stars more massive than 10~$M_\odot$ B stars, ][]{eldridge:22}.

The length ($\Delta$t) of the red and blue boxes demonstrates the timescale over which each SFR is averaged (100 and 10~Myr, respectively).  The location of the boxes along the ordinate (SFR) shows the mean SFR when measured over each of these 10 and 100 Myr timescales.   One of our conclusions is that the SED fits are able to differentiate the SFR on these timescales, and thereby place constraints on them. This impacts the interpretation of the SFR--mass relation. Before exploring the SFR--mass relation for the entire population of galaxies, we discuss how the photometric SEDs enable us to differentiate between galaxies with SFHs that are rising or declining over the past $\sim$10~Myr compared to $\sim$100 Myr timescales.
\subsection{Star Formation Rates and Inference of Spectral Features}
\label{section:sfrs}
From the SED fitting we are able to measure differences in the SFH when it is averaged over the previous $\sim$10 Myr (\sfrten) from those averaged over the previous 100 Myr (\sfrcen), using the definitions below. There are several reasons for this.   First, we have many photometric bands covering the galaxies' SEDs from the redshifted \lya\ break, to the Balmer/4000~\AA\ break.  These are sensitive to the relative number of massive stars (B-type and earlier, whose continua dominate the rest-UV) compared to less-massive stars (A-type and later) which contribute to the Balmer/4000-\AA\ break.   Secondly, these bandpasses are sensitive to the effects of emission lines in the rest-optical.  This includes primarily \oiii\ $\lambda\lambda$4960,5008+\hb, and \ha. These lines have the high equivalent widths (EWs) in star-forming galaxies \citep[i.e., ][]{smit:16, sobral:18, steidel:18, maseda:23}, and the strength appears to increase at redshifts $z\sim 2-7$ \citep[see, e.g.][]{endsley:21a, stefanon:22b, boyett:22a, endsley:23a}. Because the EWs in the observed frame increases, the emission lines impact the fluxes in the medium-- and broad--band filters to a higher degree \citep[e.g.,][]{papovich:01}. Importantly, because stellar lifetimes increase with decreasing mass, all of these features (nebular emission lines, the UV continuum, Balmer/4000~\AA\ breaks) are sensitive to the SFH on different time scales ($\sim$3--10~Myr, $\sim$ 20--100~Myr, and $\gtrsim500$~Myr, respectively). This allows us to use the multi--band dataset to constrain the SFH. 

As an example, these features are seen in the SED of CEERS\_4386 in Figure \ref{fig:soloSED} (panel b), where both \JWST/NIRCam F277W and F356W photometric bands have elevated fluxes.  Our interpretation from the SED modeling is that this is due to the presence of the \oiii +\hbeta\ and \halpha\ emission lines in these bands, respectively. For galaxies at $4.5 \lesssim z\lesssim6$, the long-wavelength NIRCam bands contain H$\alpha$, H$\beta$ and \oiii, while the shorter wavelength bands contain the rest-frame UV coverage and the 4000~\AA/Balmer breaks. For these galaxies we can place the tightest constraints on the SFH averaged over 10 and 100~Myr. For galaxies at $6 \lesssim z\lesssim9$, all of these features remain accessible to NIRCam, except for H$\alpha$, which shifts to higher wavelengths.   At $z\gtrsim 10$, we lose multi-wavelength coverage of the rest-frame optical emission lines and the $4000$\AA\ region in the NIRCam coverage,  but we retain multi-wavelength coverage of the rest-UV with the \HST/ACS imaging.  Motivated by these effects, we consider what features in the observed galaxy SEDs drive differences in the SFRs derived from the SFH when averaged over 10 Myr and 100 Myr timescales.  

Throughout, we define two quantities, \sfrten\ and \sfrcen, which are the SFR derived by averaging the SFH over 10~Myr and 100~Myr, respectively.  We then derive these for every galaxy in sample using the constraints on the SFH using:
\begin{eqnarray}\label{eq:sfrs}
    \sfrten&=& \frac{1}{10~\mathrm{Myr}} { \int_{t_\mathrm{obs}}^{t_\mathrm{obs}+10~\mathrm{Myr}}\ \mathrm{SFR}(t')\ dt'},\ \\
    \label{eq:sfrs100}
    \sfrcen&=& \frac{1}{100~\mathrm{Myr}} { \int_{t_\mathrm{obs}}^{t_\mathrm{obs}+100~\mathrm{Myr}}\ \mathrm{SFR}(t')\ dt'}, 
\end{eqnarray}
where in each case the integral spans a time from the current age of the Universe at the redshift of the galaxy ($t_\mathrm{obs}$) to the previous 10 or 100~Myr, respectively.

In Figures~\ref{fig:highsfr}--\ref{fig:lowsfr} we show examples of galaxies that have high, ``normal'' and low \sfrten\ relative to the median SFR for the galaxy sample at a given redshift.   We note that for all galaxies, there is less of a constraint at earlier ages, where all the SFH constraints allow for a possible, low-level of star-formation in the distant past.  This star-formation is certainly not excluded by the modeling as the light from the stars would be lost in the glare of the current generation of stars \citep[this is the well know ``outshining'' effect,][]{papovich:06,conroy:13}.  \citet{papovich:23} showed that even with \jwst\ data one cannot exclude these early-formed stars (and certainly \textit{some} earlier generations of stars must have existed given the propensity of emission lines from metals in the galaxies).  As we are interested in the current SFR averaged over the previous 10 and 100~Myr, possible earlier generations of stars in our galaxies do not change our conclusions here. 

Figure~\ref{fig:highsfr} shows examples of galaxies that have high \sfrten.  These galaxies all have \sfrten\ values $\approx 0.3$~dex higher than the median \sfrten\ of galaxies in the same redshift bin with $\log{(M_\ast/M_\odot)} \in [8.7,9.4]$. Each galaxy shows evidence in its photometric SED for the presence of emission lines boosting the observed flux, specifically where bands contain the redshifted \oiii+H$\beta$ and the H$\alpha$ emission lines.  The SED fits respond to these features, where the models that reproduce the data favor a recent ``upturn'' in the SFR that leads to a higher \sfrten.  This ``upturn'' in the SFR is not always sufficient to increase \sfrcen\ when the SFH is averaged over the longer timescale.  The presence of these strong emission features allows \bagpipes\ with the \textsc{dense-basis} SFHs to infer that there is a recent increase in the SFR n these galaxies.  Quantitatively, this increases the likelihood for higher \sfrten\ values compared to \sfrcen.  In the examples in Figure~\ref{fig:highsfr}, the SFR in the past 10~Myr is higher by $\log(\sfrten) - \log(\sfrcen) \geq +0.3$ dex, 1$\sigma$ above the mean SFR at fixed stellar mass and their respective redshfits, with the exception of the highest redshift bin.  

Figure \ref{fig:sfms_SED} shows examples of galaxies that have \sfrten\ within $\approx 0.3$~dex of the median \sfrten\ at their redshift and with $\log{(M_\ast/M_\odot)} \in [8.7,9.4]$.  In this sample of galaxies, there is evidence for elevated flux in the photometry due to the presence of emission lines and the blue slopes of the SEDs in the UV indicate ongoing, or very recent ($\lesssim 100$ Myr), star formation.  However, the emission line strength is generally weaker with lower EWs compared to the galaxies in Figure~\ref{fig:highsfr}.  There are also cases where the galaxy SEDs show evidence for Balmer breaks, implying the fits favor models with a larger fraction of evolved stellar populations.  Taken together, the SED fits favor models where there is little difference between \sfrten\ and \sfrcen. On average we find that $\log(\sfrten) - \log(\sfrcen) \simeq 0.0$~dex for these galaxies within their $1\sigma$ uncertainties. 

Lastly, Figure \ref{fig:lowsfr} shows examples of galaxies that have \sfrten\ $\approx 0.3$~dex lower than the median SFR of galaxies with similar redshifts and $\log{(M_\ast/M_\odot)} \in [8.7,9.4]$.  Among this sample, we observe little or no evidence for the presence of emission lines in the photometry of these galaxies.  This implies little recent star-formation in the past $\sim 10$ Myr, which is favored by the model fits as shown in the constraints on the SFHs in the Figure.  However, the SEDs \textit{do} indicate evidence for Balmer/4000\AA\ breaks, which leads to higher stellar mass--to--light ratios and lower SFRs. Taken together, the SED fits therefore favor SFHs where the \sfrten\ is lower than \sfrcen.  For these galaxies we find $\log(\sfrten) - \log(\sfrcen) \leq -0.3$~dex, $1\sigma$ below the mean \sfrten at fixed mass and redshift.

Therefore, the SED fitting is able to provide meaningful constraints on the SFH over the past 10 and 100~Myr.  This information is important for interpreting the evolution of the SFMS at these redshifts.   This is primarily true for galaxies at $4.5 < z < 9$, where (as mentioned above) the NIRCam photometry samples the full SED from the rest-frame UV through the rest-frame optical, capturing the UV continuum,  important nebular emission lines in the rest-optical, and the region around the Balmer/4000~\AA\ break.  In contrast, at $z > 9$, we lose the coverage of these redder features, which negatively impacts the constraints we can measure on the SFHs.  We therefore primarily focus on the evolution of the SFRs and stellar masses at $4.5 \leq z \leq 9$.  We retain the galaxies in the $9 < z <12$ sample in our analysis as this provides hints for this population, but we acknowledge that additional data (e.g., MIRI imaging) are needed to cover the same rest-frame optical features as the $z \leq 9$ sample. 
\section{Results}
\label{section:results}
Using the SFR and stellar-mass constraints, we now focus on the SFR--$M_\ast$ plane to study the SFMS for the galaxies at $4.5 < z < 12$. Figure \ref{fig:sfrm_full} shows the SFR--$M_\ast$ plane for all galaxies in our sample.  The panels show the relation using the SFRs averaged over the previous 10 Myr (left panel, \sfrten) and 100 Myr (right panel, \sfrcen, as defined in Equations \ref{eq:sfrs} and \ref{eq:sfrs100}, respectively). One immediate conclusion from these figures is that the scatter in the SFR--$M_\ast$ (in $\log(\mathrm{SFR})$)  increases dramatically when the SFRs are averaged over the shorter 10~Myr timescale.  However, there is little evolution in either the slope or normalization with redshift in the SFMS.

It is important to note that we make no selection in specific SFR in our analysis of the SFR--$M_\ast$ relation.  However, all galaxies in our sample would be considered star-forming when using \sfrcen, as none would qualify as ``quiescent'' galaxies. The lowest s\sfrcen\ in the sample is log(s\sfrcen/yr$^{-1}$)~$\simeq~-9.5$, above the canonical cut-off for quiescent galaxies at $z \leq 6$.  For example, \citet{tacchella:22} advocate for a sSFR selection of quiescent galaxies that can be as high as $(\mathrm{sSFR}/\mathrm{yr}^{-1})\simeq~-10$. Similarly, other studies \citep[see e.g.,][and references therein]{estrada-carpenter:23} have shown that the sSFR separation between star-forming and quiescent galaxies slowly increases to higher sSFRs at higher redshifts but even extrapolating these trends to the highest redshifts in our sample implies that all the galaxies in our sample are ``star-forming'' when considering s\sfrcen.  We therefore make no distinction between galaxies that are star-forming or quiescent in our sample.   
%
%
\subsection{The SFMS in the Epoch of Reionization}
\label{section:sfms}
\begin{figure*}[t!]
    \centering
    \includegraphics[width=\linewidth]{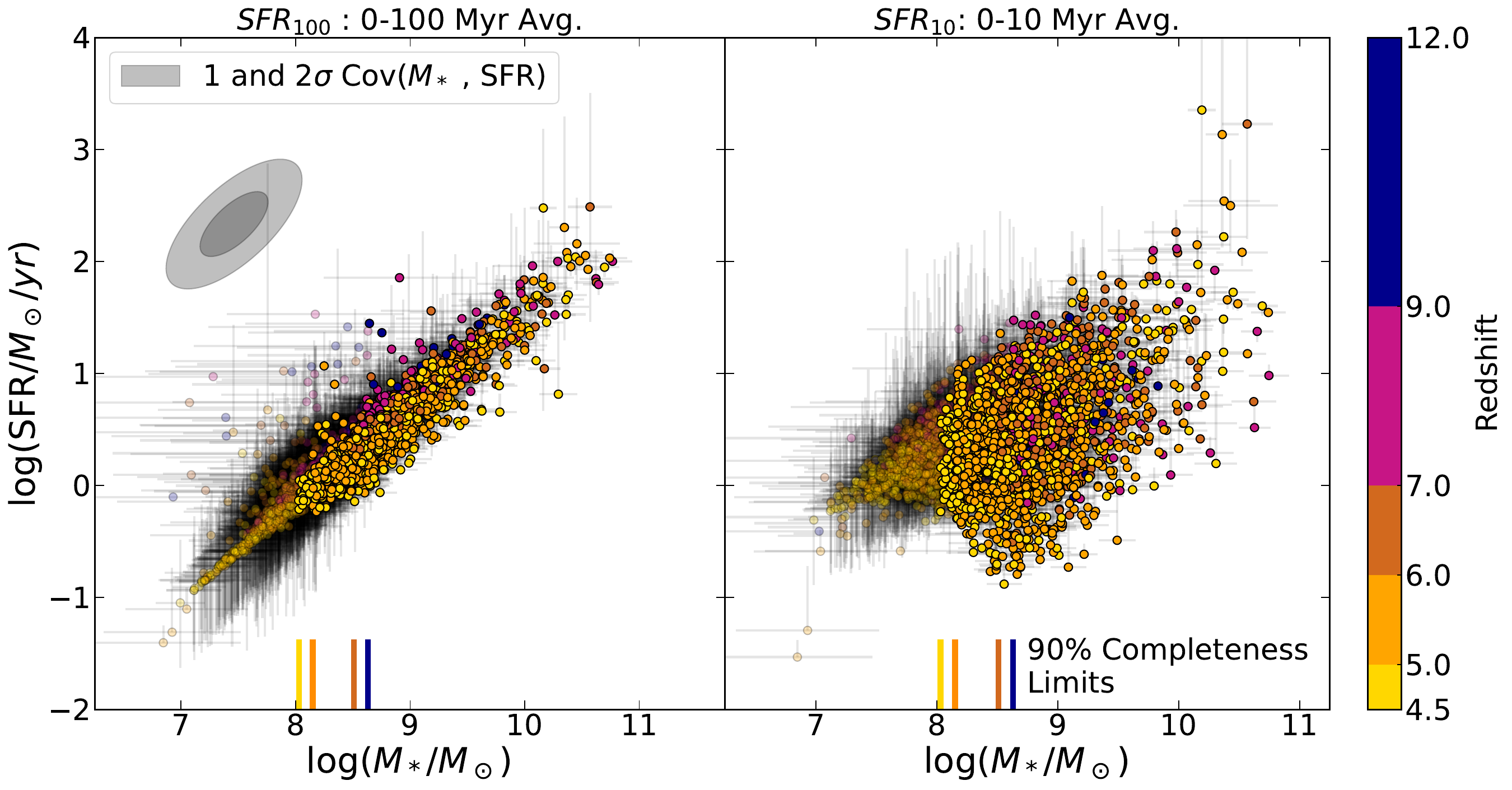}
    \caption{The SFR--M$_\ast$ relation illustrated as the star-forming main sequence (SFMS) for the full sample of galaxies above our estimated stellar mass completeness limits. Data points show these galaxies, color-coded by redshift. Faded points show galaxies below the 90\% completeness limits at their respective redshifts, which are indicated by the colored, vertical lines at the bottom of each panel. The error bars indicate the 68\% uncertainty on each point. The uncertainties on mass and SFR are covariant, and the gray shaded ellipses in the left panel shows a representative 1$\sigma$ and 2$\sigma$ posterior ellipse in SFR--M$_\ast$ space for a single galaxy (the size of the ellipse may change, but the orientation of the covariance is similar for all galaxies in both \sfrten\ and \sfrcen).  In the left panel, we show the SFR--M$_\ast$ relation using the star formation rates measured over 100 Myr (\sfrcen) and in the right panel, we show the same but with star formation rates measured over 10 Myr (\sfrten).  We observe little evolution in either the slope or normalization with redshift in the SFMS, but the scatter in the SFMS increases when using the \sfrten\ values.  }
    \label{fig:sfrm_full}
\end{figure*}

We parameterize the SFMS as a power-law function between the SFR and stellar mass, $M_\ast$, where $\log \mathrm{SFR} \sim \alpha \log M_\ast$.  We divided our sample into five bins of redshift (listed in Table~\ref{table:MCMC_results}). Two of these bins lie at $z < 6$ \citep[nominally after the end of the epoch of reionization, ][]{becker:21}, and two bins at $6 < z < 9$ (during reionization).  We also include a bin with $9 < z < 12$, but we note that at these redshifts the measurements of the \sfrten\ for galaxies are less-well constrained as the rest-frame optical nebular emission lines shift out of the NIRCam wavelength coverage. 

To account for mass--completeness, we include only galaxies in each redshift bin that have stellar masses above an estimated 90\% completeness limit. We derived the latter following the method described in \citet{marchesini:09}.  We take all galaxies with $m_\mathrm{F277W} \leq 27.7$~mag, 0.5 mag above our selection limit.  We dim these galaxies (in magnitude) and lower their stellar mass by a corresponding amount in small increments until we reach the magnitude ($m_\mathrm{F277W}$) at which 90\% of the galaxies are still brighter than our selection criteria at a given redshift.   We then take the stellar mass corresponding to the brightest galaxy that would fail our magnitude limit ($m_\mathrm{F277W} \leq 28.2$~mag) as the mass--completeness limit. We do this for each redshift bin as the stellar mass threshold evolves with redshift.  We list the stellar-mass completeness limits in Table~\ref{table:MCMC_results}. 
%

To model the SFMS, we define a power-law relation between the SFR and stellar mass as,
\begin{equation}
    \log(\mathrm{\Psi_\ast}) = \alpha \log\left(\frac{M_{*}}{10^{8.74}~M_\odot}\right) + \beta(z) + \epsilon(z),
    \label{eq:loglinear}
\end{equation}
where $\Psi_{*}$ is the SFR in solar masses per year, and $\alpha$ is the slope of the SFMS.  There are three parameters, $\alpha$, $\beta(z)$, and $\epsilon(z)$.  Here, we assume a single value for the slope, $\alpha$, that does not evolve with redshift over $4.5 < z < 12$.  This is an assumption supported by the fact that we observe no evidence in a change in slope from Figure~\ref{fig:sfrm_full}, particularly in the $\sfrcen$--$M_\ast$ relation.  However, this may be a result of the limited dynamic range we have in stellar mass in our sample.  Larger-area and deeper datasets will be needed to provide better constraints on the evolution of this slope than what we can achieve here.   We then allow for evolution in the normalization of the relation, $\beta(z)$, here defined at a mass of $\log (M_\ast/M_\odot) = 8.74$, the median stellar mass of our sample.  

The variable, $\epsilon(z)$, allows for the intrinsic scatter along the relation.  We assume $\epsilon(z)$ is drawn from a normal distribution with mean, $\mu=0$ and variance = $\sigma_\mathrm{int}^2$, that is, $\epsilon(z) \sim N(0,\sigma_\mathrm{int}(z)^2)$.  The value of $\sigma_\mathrm{int}(z)$ then represents the intrinsic scatter in the SFMS at each redshift (assuming here no dependence on stellar mass, but we consider such a dependence below). 

We use a Monte Carlo sampler to determine the best fit and uncertainties.    We use the Markov Chain Monte Carlo (MCMC) routine implemented with the Python package \textsc{emcee} \citep{foreman-mackey:13} for this purpose.   In the MCMC fit, for every walker in each iteration, we fit a single $\alpha$ for all galaxies (in all redshift bins), where there is a separate $\beta(z)$ in each redshift bin.   In each  iteration of the fit, for each galaxy, we resample from the covariance between the variables in the fit (i.e., $P(M_\ast, \mathrm{SFR})$). That is, we draw a new SFR \textit{and} stellar mass from the 2-dimensional posterior space for each galaxy, and in this way galaxies are allowed to move within their respective posterior spaces in each iteration of the MCMC fit.  This is important as the SFR and stellar mass are strongly correlated (see gray ellipses in Figures~\ref{fig:sfrm_full}--\ref{fig:sfrm_split_100}).

The details of the likelihood used in the fit are provided in Appendix~\ref{section:appendix}.  In summary, we split the likelihood into two parts.   First, we model the relation, 
\begin{equation}\label{eq:noscatter}
    \log{(\Psi_{*})} = \alpha \log\left(\frac{M_{*}}{10^{9}~M_\odot}\right) + \beta(z),
\end{equation}
which becomes the first part of the likelihood function. Second, we  subtract this relation from from the observed data to model the scatter about the relation, $\epsilon(z)$.  We then fit this in each redshift bin, taking the absolute value of the scatter and use a $\chi^2$-squared statistic to determine the likelihood, assuming Gaussian scatter about the SFMS, to compare to the absolute value of a normally distributed sample with the walker's associated variance. The $\chi^2$-squared likelihood from this step is the second part of our likelihood function. We provide tests and further descriptions of this methodology in Appendix~\ref{section:appendix} and show that we are able to recover accurately the slope, normalization, and scatter in the relation.

We fit the SFMS separately for the \sfrten\ and \sfrcen\ values.   Figures \ref{fig:sfrm_split_100} and \ref{fig:sfrm_split} show the results for the SFMS from the MCMC fits.  Table~\ref{table:MCMC_results} lists the measured values for each parameter.  
We discuss the results in the next subsection.

\begin{figure*}[t!]
    \centering
    \includegraphics[width=\linewidth]{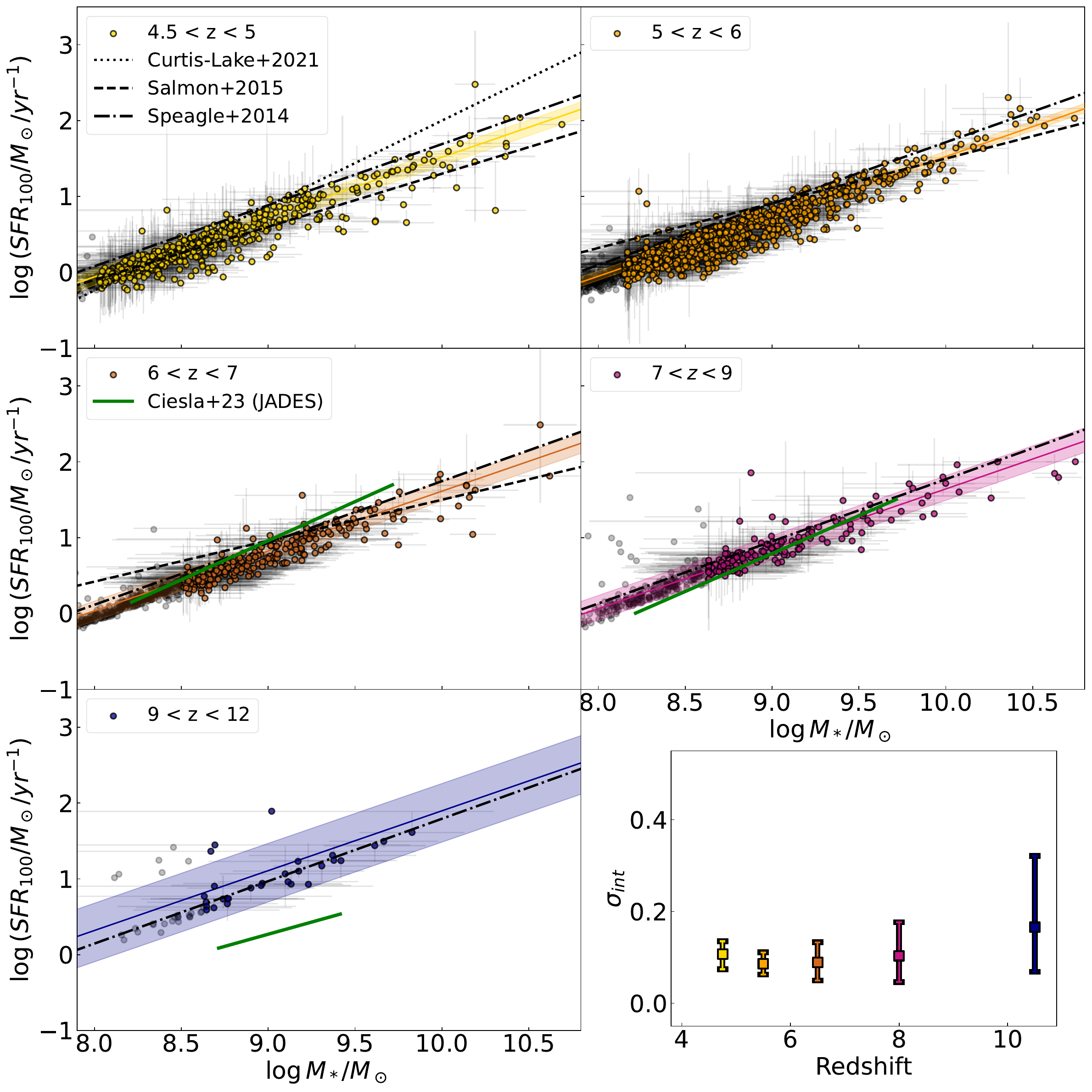}
    \caption{The SFR - $M_\ast$ relation for the full sample of galaxies using the SFRs averaged over the past 100~Myr (\sfrcen).  Each panel show the relation for a different redshift bin, as labeled. The data points show median values for each galaxy where the error bars show the inner-$68^{th}$ percentile of the posteriors on \sfrcen\ and $M_\ast$.  Dark-filled data points lie above the 90\% mass completeness limit (fainter gray symbols lie below this limit and are excluded from the fit).   The solid, colored lines show the median best-fit relation from the MCMC fit (Equation~\ref{eq:noscatter}), where the shaded region shows the $1\sigma$ error on this relation. For comparison, we show measurements from the literature at $z \lesssim 7$ \citep{curtis-lake:21, salmon:15} and we extrapolate the relation from \citet{speagle:14} assuming that their relation continues its evolution out to $z=12$. The solid-green line shows the measurements at $z \gtrsim 6$ from the JADES survey \citep{ciesla:23}.  The bottom-right-most panel shows the inferred intrinsic scatter, $\sigma_{int}$, as a function of redshift with asymmetric errors estimated from the MCMC posterior. }
    \label{fig:sfrm_split_100}
\end{figure*}

\begin{figure*}[t!]
    \centering
    \includegraphics[width=\linewidth]{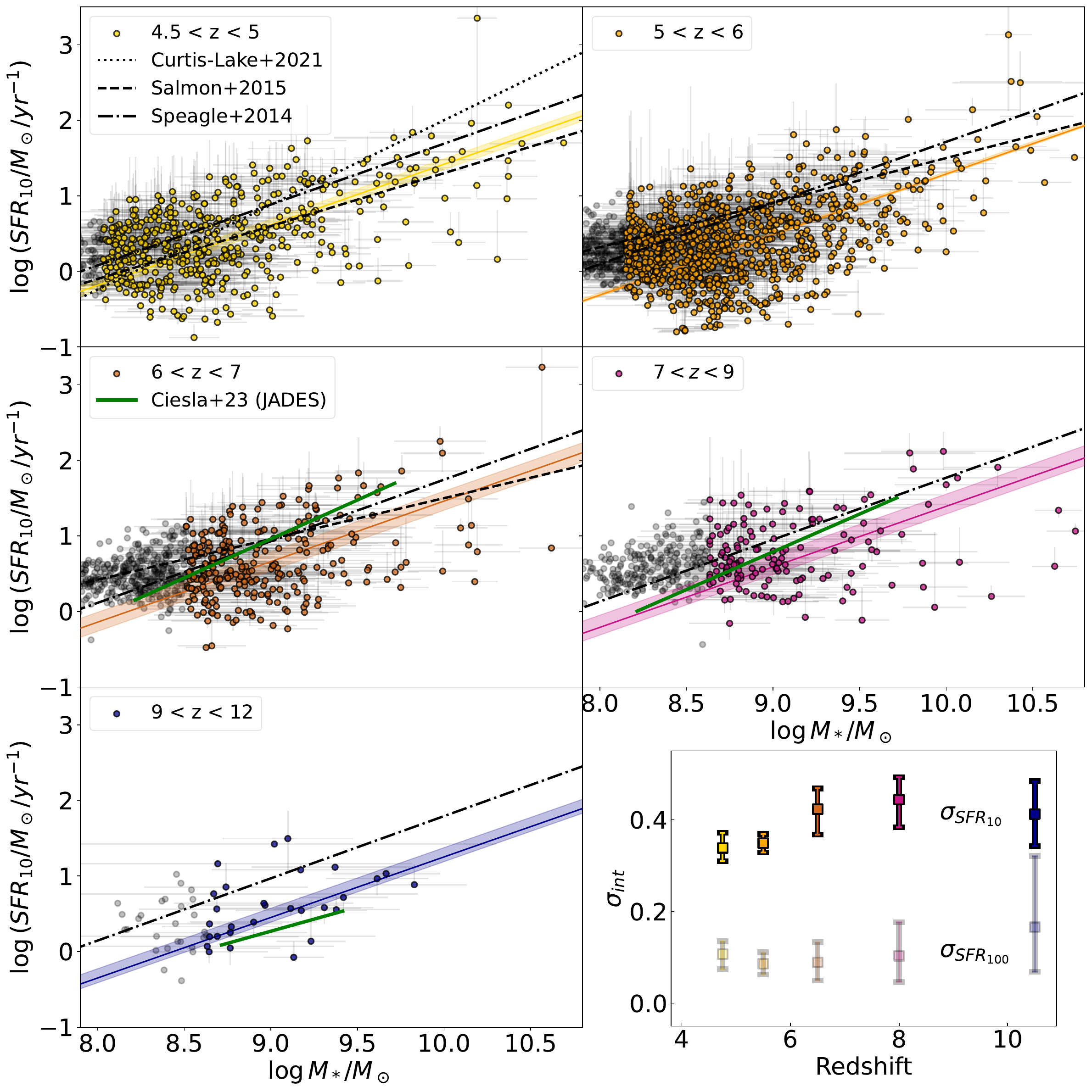}
    \caption{The same as Figure \ref{fig:sfrm_split_100}, but the SFR averaged over the previous 10~Myr (\sfrten).   In the bottom-right-most panel, we show the results for $\sigma_{int}$ for \sfrten\ as darker points.  For comparison, the lighter points show the $\sigma_{int}$ measurements from \sfrcen\ (from Figure~\ref{fig:sfrm_split_100}). The scatter derived from the \sfrten\ measurements are substantially higher than those derived from \sfrcen.}
    \label{fig:sfrm_split}
\end{figure*}

\begin{deluxetable*}{ccccccc}
\caption{Results from the MCMC analysis for both SFR$_{10}$ (top) and SFR$_{100}$ (bottom).}
\tablehead{
\colhead{Redshift} & \colhead{Sample Size} & \colhead{\begin{tabular}{c} stellar mass\\[-2pt] 90\% Completeness\end{tabular}} & \colhead{$\alpha$} & \colhead{Median~$\log{(M_\ast)}$} & \colhead{$\beta(z)$} & \colhead{$\sigma(z)$}
\label{table:MCMC_results}}
\startdata
    \sidehead{MCMC Results Measured With SFR$_{100}$}
    \hline
    All galaxies & 1863 & & $0.79\pm0.04$ & $8.74$ & & \\
    \hline
    $4.5 < z \leq 5$ & 475 & 8.03 & & & ${0.51}\pm0.05$ & ${0.10}\pm0.03$ \\
    $5 < z \leq 6$ & 921 & 8.15 & & & ${0.52}\pm0.02$ & ${0.09}\pm0.02$ \\
    $6 < z \leq 7$ & 291 & 8.51 & & & ${0.61}\pm0.06$ & ${0.09}\pm0.05$ \\
    $7 < z \leq 9$ & 147 & 8.63 & & & ${0.64}\pm0.07$ & ${0.10}\pm0.07$ \\
    $9 < z \leq 12$ & 29 & 8.63 & & & ${0.89}\pm0.05$ & ${0.17}\pm0.15$ \\  
    \hline
    \sidehead{MCMC Results Measured With SFR$_{10}$}
    \hline
    All galaxies & 1863 & & $0.80\pm0.01$ & $8.74$ & & \\
    \hline
    $4.5 < z \leq 5$ & 475 & 8.03 & & & ${0.40}\pm0.04$ & ${0.34}\pm0.03$ \\
    $5 < z \leq 6$ & 921 & 8.15 & & & ${0.27}\pm0.02$ & ${0.35}\pm0.02$ \\
    $6 < z \leq 7$ & 291 & 8.51 & & & ${0.44}\pm0.06$ & ${0.42}\pm0.05$ \\
    $7 < z \leq 9$ & 147 & 8.63 & & & ${0.37}\pm0.07$ & ${0.44}\pm0.06$ \\
    $9 < z \leq 12$ & 29 & 8.63 & & & ${0.23}\pm0.05$ & ${0.41}\pm0.08$ \\ 
\enddata
\tablecomments{Fitted parameters for Equation \ref{eq:loglinear} are shown in bold ($\beta(z)$ and $\sigma^2(z)$) and we list $\alpha$ and the median stellar mass, $\left< M_* \right>$ for clarity.}
\end{deluxetable*}

\subsection{The Slope and Normalization of the SFMS Relation}

Figure \ref{fig:sfrm_split_100} shows the results of the SFMS fits with \sfrcen. 
The first five panels (left-to-right, top-to-bottom) of Figure~\ref{fig:sfrm_split_100} show the results for each of the five redshift bins ($4.5 < z < 12$), as labeled. The bottom-right panel shows the results for $\sigma_{\mathrm{int}}(z)$, or the intrinsic scatter along the SFMS. 

The SFMS we derive using the \sfrcen\ values is consistent with those from previous measurements at these redshifts \citep[all measured using SFRs sensitive to the SFH over 100 Myr timescales, ][]{speagle:14, salmon:15, curtis-lake:21}.  The \citet{speagle:14} relation is measured at lower redshifts.  However, if we extrapolate this to the redshifts and values for our sample, the slope and normalization of the SFMS we derive agree.   Similarly, our results also agree  with those from \citet{salmon:15}, although they find evidence that the slope of the SFMS flattens with increasing redshift.  Here, we assume a constant slope, where the evolution is then in the normalization.  The combination of these yields a similar SFMS relation (see Figure~\ref{fig:sfrm_split_100}).

Figure \ref{fig:sfrm_split} shows the SFMS fits using SFRs over 10~Myr timescales.  The results from these fits are listed in the bottom rows of Table \ref{table:MCMC_results}. The bottom right panel of the figure again shows the intrinsic scatter, $\sigma_\mathrm{int}(z)$, using \sfrten\ (color-coded points and error bars), and the faded points in this panel show the $\sigma_\mathrm{int}(z)$ values derived using the \sfrcen\ shown in Figure \ref{fig:sfrm_split_100}. Using \sfrten, we measure a slope for the SFMS of $\alpha=0.80\pm0.01$, consistent with our measurement of the SFMS when using \sfrcen.  However, the normalization values for $\beta(z)$ are lower when using \sfrten\ compared to \sfrcen.  This implies that \textit{on average} the \sfrten\ values are lower than \sfrcen\ (at fixed stellar mass).  This is consistent with our measurement of the duty cycle for star-formation (discussed below in Section \ref{section:DC}). 

We again find that the SFMS we derive using \sfrten\ is consistent with values from the literature.  The most relevant comparison is to the work of \citet{smit:16} who used broad-band photometry from \textit{Spitzer}/IRAC to estimate the \ha\ fluxes (and thus SFRs) for a spectroscopic sample of massive star-forming galaxies at $\langle z \rangle = 4.25$.  As the \ha\ flux traces the most massive stars it also will respond to variations on 10~Myr timescales, and thus similar to our \sfrten\ values \citep{kennicutt:98}.     \citet{smit:16} derive a slope $\alpha=0.78\pm 0.23$, well matched to ours from the SFMS using \sfrten, $\alpha = 0.79\pm0.04$.  Similarly, they derive a normalization for galaxies at $\log M_\ast/M_\odot=10$ of $\beta = \log \mathrm{SFR} / (M_\odot~\mathrm{yr}^{-1}$ = $1.51 \pm 0.07$.  If we extrapolate our normalization to this mass we obtain $\beta(z=4.5)$ = $1.40 \pm 0.04$, consistent at the $1\sigma$ level.   This suggests some consensus in the shape of the SFMS at $z\sim 4-5$ when using galaxy SFRs averaged over 10~Myr.  However, the \textit{scatter} in the relation has yet to reach consensus (which we discuss in the next subsection). 

The evolution in the normalization of the SFMS depends on the choice of SFR averaged over the different timescales.  Using \sfrcen\ we measure a slow increase in normalization with increasing redshift, going from $\beta \simeq 0.5$ at $z=4-6$ to $\beta \simeq 0.6$ at $z=6-9$.    This agrees with previous studies at $1\leq z \leq6$ \citep[e.g, ][]{whitaker:12, speagle:14, salmon:15, popesso:23}.  However, this trend is markedly different for the normalization of the SFMS using the \sfrten\ values.  These are consistently lower than the normalization values derived using \sfrcen, and show no evidence of evolution from $z=4.5$ to 9.  We do not consider the trend to higher redshifts, $z > 9$, to be significant as at these redshifts the strong optical emission lines tracing the SFR on 10~Myr timescales shift beyond the coverage of the \jwst/NIRCam data. This has implications for the duty cycle (because on average the \sfrten\ values are smaller than \sfrcen), and we consider this in Section~\ref{section:DC} below.

We see possible evidence that the normalization of the SFMS using \sfrten\  is lower for galaxies in the $5 < z < 6$ redshift bin.  This is notable as there is a purported galaxy overdensity in CEERS in this bin, at $z\simeq 5.1$ \citep{chworowsky:23a}, and it may be that this modifies the normalization of the SFMS relation through environmental processes, similar to other results at lower redshifts, $z\sim0$ to $z\sim 1$ \citep{old:20,boselli:23}.  It will be important to compare the results here with similar analyses in other fields to test if effects of the large-scale structure are biasing our conclusions. 
\subsection{Scatter in the SFMS Relation}
\label{section:sfms_scatter}
As illustrated in Figure~\ref{fig:sfrm_split} (lower-right panel), when using \sfrten, there is an apparent increase in the intrinsic scatter in the SFMS, $\sigma_\mathrm{int}(z)$, with increasing redshift from $z\sim4.5$ to $z\sim 9$.  Fitting to the full samples (over all stellar masses) we see the intrinsic scatter increases from $\sigma_\mathrm{int} = 0.34\pm 0.03$~dex at $4.5 < z < 5$ to $\sigma_\mathrm{int} = 0.44 \pm 0.06$~dex at $7 < z < 9$.  In contrast, there is no evidence for evolution in $\sigma_\mathrm{int}$ when using the SFRs averaged over the longer 100~Myr timescales, \sfrcen.   The interpretation is that there is an increase in the variability on the shorter timescale of \sfrten\ for galaxies at higher redshifts.  However, these values for $\sigma_\mathrm{int}$ are measured for the entire galaxy population at each redshift.  

\begin{figure*}
    \centering
    \includegraphics[width=0.8\linewidth]{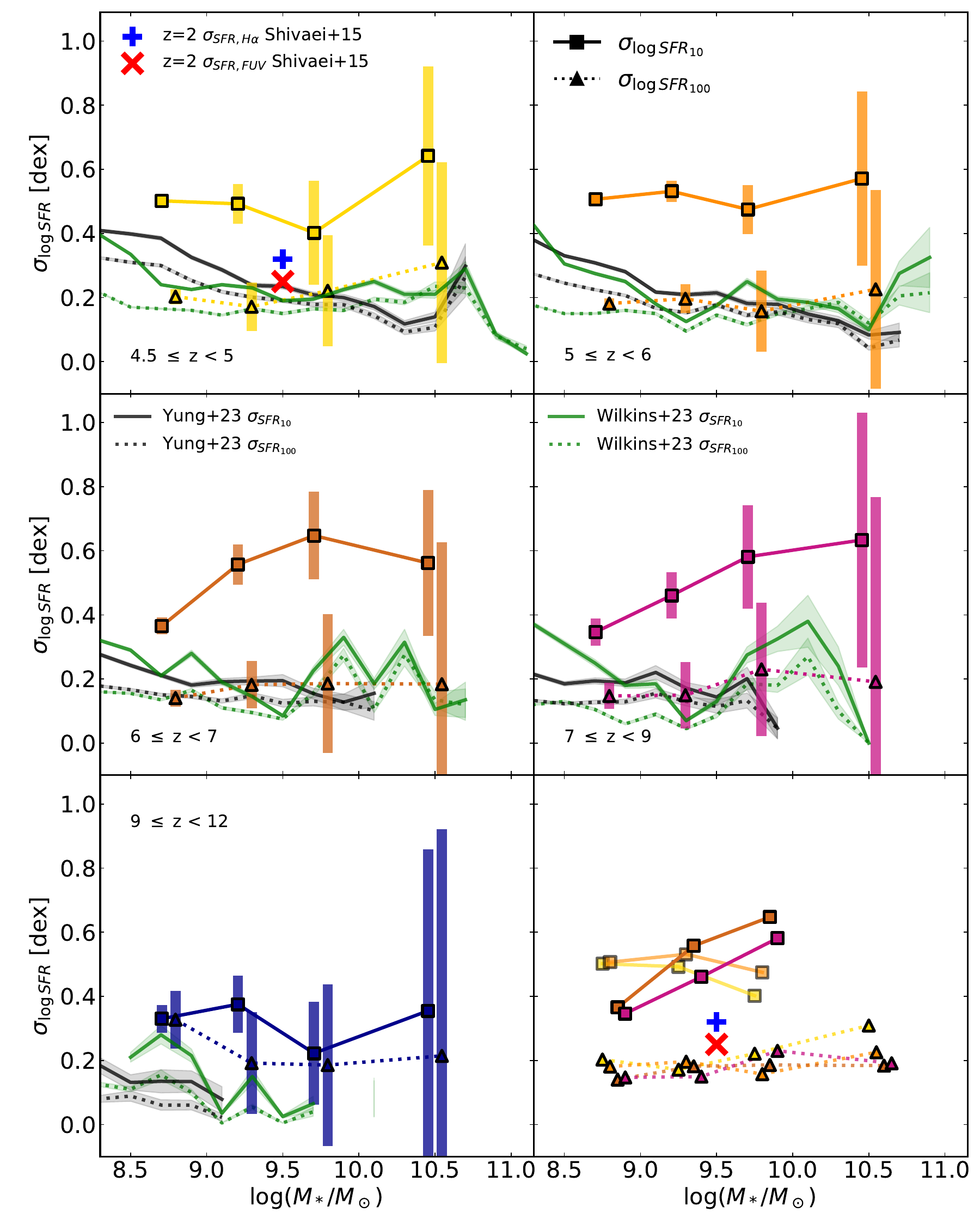}
    \caption{Observed scatter in the SFMS as a function of stellar mass and redshift. Each panel shows a different redshift, where the colors denote different redshift bins as in Figure \ref{fig:sfrm_split} and as labeled. Square markers connected by solid lines show the observed $1-\sigma$ scatter for \sfrten\ (the SFR averaged over 10~Myr) and triangles connected by dotted lines show the observed $1-\sigma$ scatter in \sfrcen\ (the SFR averaged over 100~Myr). Error bars on $\sigma_\mathrm{log SFR}$ are estimated using equation \ref{eq:sigma_err}. In the panel for each redshift, we show theoretical predictions for $\sigma_{\log{(\mathrm{SFR})}}$ from the SC-SAM \citep[][black]{yung:23b} and FLARES \citep[][green]{lovell:21, wilkins:23b}, where solid (dotted) lines show these predictions for \sfrten\ (\sfrcen) and the shaded regions indicate the 84$^{th}$--to--16$^{th}$ percentiles.  In the top-left panel (and bottom-right-most panel), the X and cross show the measured observed scatter along the SFMS at z=2 from \citet{shivaei:15} for FUV star formation rates and \halpha\ star formation rates (as labeled). The bottom-right-most panel compares the results for all redshift and stellar mass bins  (error bars are removed, and points at the highest stellar masses are removed for clarity).   }
    \label{fig:sfrcomp}
\end{figure*}

We also measured the evolution in the intrinsic scatter along the SFMS as a function of stellar mass in each redshift bin.    We perform this measurement using both \sfrten\ ($\sigmaten$) and \sfrcen\ ($\sigmacen$). For each stellar mass and redshift bin, we recalculated the scatter as half the distance from the $16^{th}$ to $84^{th}$ percentile (i.e., the inner 68$^{th}$-percentile) of the distribution in SFR. We again assume that SFRs are distributed normally, and therefore, half of the inner $68^{th}$-percentile is defined as $\sigma$ and we can estimate the uncertainty on the intrinsic scatter as,
\begin{equation}
    \label{eq:sigma_err}
    \delta\sigma = \frac{\sigma}{\sqrt{2(N-1)}},
\end{equation}
where $N$ is the number of galaxies in the estimation of $\sigma$.
We show these measurements of the scatter and their uncertainties in Figure \ref{fig:sfrcomp}.  We include as a comparison measurements from the literature \citep{shivaei:15} and predictions from models and simulations \citep{yung:23b,wilkins:23b}.

There are several interesting trends in Figure~\ref{fig:sfrcomp}. First, $\sigmacen$ is roughly constant in both stellar mass and redshift over the entire range of our sample, with $\sigmacen \approx 0.2-0.3$~dex.  This is consistent with previous results from lower redshifts, within the quoted uncertainties.  \citet{shivaei:15} find that the scatter in SFR derived for galaxies at $z\sim 2$ from rest-FUV data is $\sigma_{\mathrm{FUV}} \simeq 0.3$~dex.  \citet{schreiber:15} obtain similar results, finding $\sigma \simeq 0.3$~dex using far-IR data from \textit{Herschel} for massive star-forming galaxies at $2 < z < 4$ with no apparent evolution in redshift nor stellar mass.   As these observables probe star-formation on $\sim100$~Myr timescales, this is consistent with our observed scatter in $\sfrcen$. \citet{ciesla:23} recently used \jwst/NIRCam data to measure the scatter in the UV absolute magnitudes of galaxies at $z > 6$, finding $\sigma_\mathrm{UV}\simeq 1.2$~mag.  Converting this to a scatter on the SFR yields $\sigma_\mathrm{UV} \simeq 0.5$~dex, but this includes both the scatter in the SFR and  the dispersion in dust attenuation.    As the latter is typically $\sigma(A_\mathrm{FUV})$=0.5-0.6~mag \citep{alvarez-marquez:16} this may account for at least half of the scatter in  $\sigma_\mathrm{UV}$.  We therefore consider these measurements consistent with our value for \sigmacen\ here. 

Second, at all redshifts \sigmaten\ is higher than \sigmacen, with typical values of $\sigmaten\simeq 0.4-0.6$~dex.  This is illustrated in the bottom-right panel of Figure~\ref{fig:sfrcomp}. At lower redshifts, $4.5<z<6$, \sigmaten\ shows no dependence on stellar mass, with a roughly constant value of $\sigmaten \simeq 0.5$~dex.  At higher redshifts, $6<z<9$, there is a positive correlation between stellar mass and \sigmaten:  the scatter increases with stellar mass, from $\sigmaten \simeq 0.4$~dex at $\log M_\ast/M_\odot \simeq 8.75$ to $\simeq 0.6$~dex at $\log M_\ast/M_\odot = 10$.  

Comparing to measurements of $\sigma_\mathrm{SFR}$ using H$\alpha$ data at $z\sim 2$ from \citet{shivaei:15}, the observed scatter in \sfrten\ we obtain is  higher by a factor of $\sim$2.  This may imply that there is an increase in the scatter when averaged over $\sim10$~Myr timescales from $z\sim 2$ to $z > 4$ such that SFHs contain higher variability at higher redshifts.    It is also possible that studies that require detections in \ha, could lead to a bias toward lower scatter (a form of Malmquist bias).   \citet{shivaei:15} included only galaxies with \ha\ detections where our sample includes all galaxies, including objects that lack evidence for strong nebular emission in their photometry (see Figure~\ref{fig:lowsfr}).   The presence of these lower--SFR galaxies increases the scatter in our measurement.   Confirming the evolution in \sfrten\ could be tested with spectroscopy around the regions of the strong rest-optical emission lines (e.g., Balmer lines, \oiii, etc) from future \jwst\ spectroscopy for galaxies at $z > 4.5$.  

At the higher redshifts, $6 < z < 9$, the scatter in \sfrten\ appears to increase with increasing stellar mass ($\log M_\ast/M_\odot \gtrsim 9$).  However, we consider this may be a systematic or selection effect. It is arguable that our samples are less complete in SFR at these stellar masses in the highest redshift bins.   We have estimated this effect to ensure the data are complete in stellar mass at 90\% (see Section~\ref{section:sfms}). However this estimate assume a slowly changing population in both luminosity and stellar mass. If there is substantially more scatter in the SFR at lower stellar masses, then the SFR distribution may be truncated and biased away from galaxies with lower SFRs.  This would make our estimate incomplete and artificially tighten the distribution (another instance of Malmquist bias). This can be tested with deeper imaging data from \jwst. However, the stellar-mass completeness will not impact the higher-mass end of the SFMS, $\log M_\ast/M_\odot \gtrsim 9.5$, which is where we observe a significant increase in the observed scatter from $4.5<z<5$ to $7<z<9$.  We can identify no systematic effect that would impact this trend, which we therefore consider to be robust. 

In Figure \ref{fig:sfrcomp}, we also compare to predictions for the scatter in the SFMS from the Santa Cruz semi-analytic model \citep[SC SAM;][]{somerville:15, yung:19a} and FLARES simulation \citep{lovell:21}.  We will return discuss the comparison between our results and these predictions in Section~\ref{section:discussion:models}.

\subsection{The Distribution of specific SFR}

\begin{figure}[t!]
    \centering
    \includegraphics[width=\linewidth]{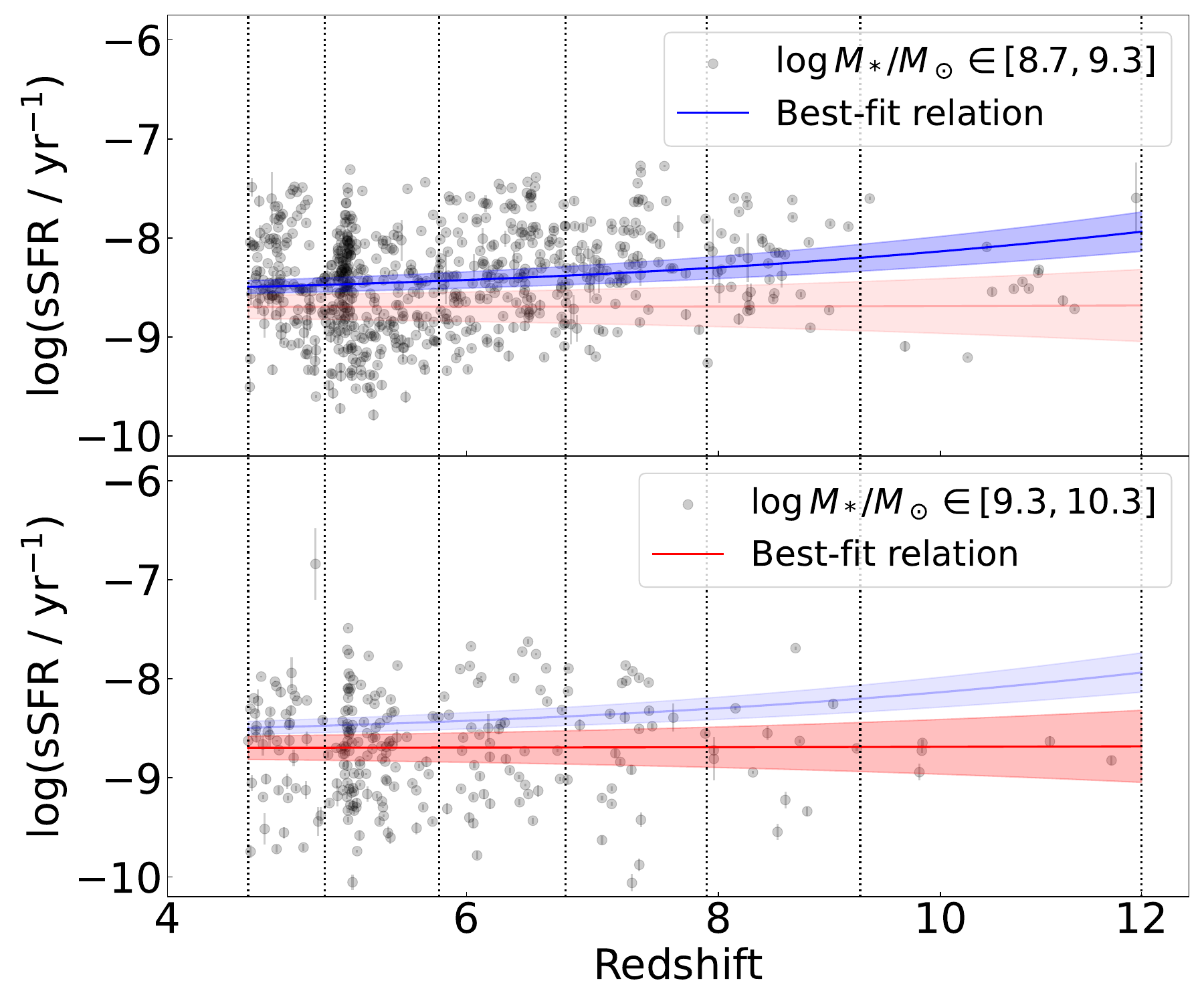}
    \caption{Specific SFRs (sSFRs) measured from \sfrten\ as a function of redshift for  the galaxies in our sample in bins of stellar mass. The top panel shows lower-mass galaxies with $10^{8.7} \leq M_*/M_\odot < 10^{9.3}$.  The bottom panel shows higher-mass galaxies with $10^{9.3} \leq M_*/M_\odot < 10^{10.3}$.  The gray circles in both panels show the mean log(s\sfrten) for individual galaxies in each sample, where the error bars represent the inner $68^{th}$-percentiles (these includes the contribution from the covariance between stellar-mass and SFR, see Section \ref{section:sfms}). Vertical dotted lines show even lookback-time steps of 200 Myr to demonstrate the trends in both redshift and cosmic time. The blue and red lines and shaded areas show the best-fit and $1\sigma$ range from parabolic fits to the data top (lower-mass galaxies) and bottom panels (higher-mass galaxies), respectively.  The curves for each bin of stellar mass are shown in both panels to contrast the results (but are dimmed for clarity), as labeled.   At all redshifts the higher-mass galaxies have lower s\sfrten, where the offset is consistently 0.3~dex.  This is significant at the $>1.5\sigma$ level at all the redshifts of our sample. }
    \label{fig:ssfr_z}
\end{figure}

Given the apparent differences in the observed scatter in galaxy SFRs with stellar mass, we test whether these trends are also apparent in the distributions of specific SFR (sSFR). Figure \ref{fig:ssfr_z} shows the s\sfrten\ (the specific SFR using the SFRs averaged over 10~Myr timescales) as a funciton of redshift in two bins of stellar mass, $\log M_\ast/M_\odot \in [8.7,~9.3]$, and $\in [9.3,~10.3]$.   To study the trends, we fit a parabolic function to the data in each bin of stellar mass allowing for evolution in redshift, $z$, following the form,
\begin{equation}
    \log{(\mathrm{sSFR}(z))} = A \times z^2 + B.
    \label{eq:ssfr}
\end{equation}
We measure the parameters using a simple least-squares regression.  The fitted models for each mass bin are shown in both panels of Figure~\ref{fig:ssfr_z} for comparison, where we show the median best-fits as the solid lines and 1-$\sigma$ range as the lighter shaded regions.  At all redshifts, the median s\sfrten\ for higher-mass galaxies is lower by $\Delta$log(s\sfrten) $\approx$0.3~dex than those for lower-mass galaxies. This difference is significant at the $>1.5\sigma$ level at all redshifts of the galaxies in our sample. 

In addition, the scatter in the distribution of s\sfrten\ is larger for the higher mass galaxies. For the higher mass galaxies, we measure $\sigma(\log\mathrm{s\sfrten})$ =  0.35 dex at $z\sim 5$ and 0.45 dex at $z\sim 8$.  For the lower-mass galaxies, we find $\sigma( \log \mathrm{s\sfrten})$ = 0.36 and 0.35 at $z\sim 5$ and 8, respectively.  In all cases this is consistent with our observation above (Figure \ref{fig:sfrcomp}) that the scatter in the \sfrten\ is increasing with increasing mass and redshift.  Here we find this is also the case when using the specific SFRs. 

\begin{figure}
    \centering
    \includegraphics[width=\linewidth]{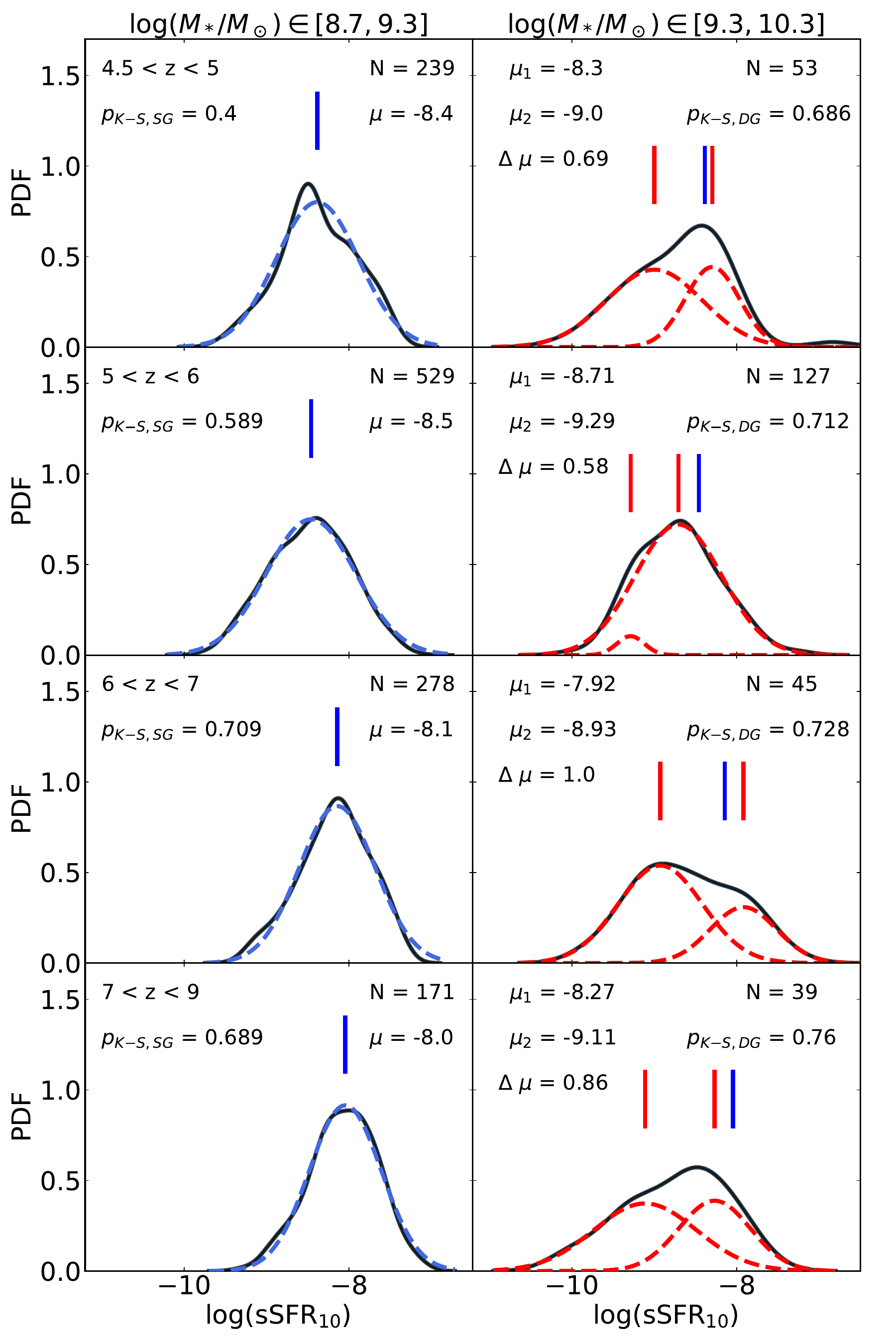}
    \caption{ Specific \sfrten\ (s\sfrten) distributions for our galaxy sample in bins of mass and redshift. In each panel of the left column we show a kernel density estimate (KDE) for the lower-mass galaxies with  log($M_*/M_\odot$) $<$ 9.3 as the black, solid line.  A best-fit Gaussian model fit to the distribution is shown as the blue, dashed line. In each panel of the right column we show a KDE for the higher-mass galaxies, $\log(M_\ast/M_\odot) \in [9.3, 10.3]$ as the black, solid line.  A best-fit double-Gaussian model is shown as the red, dashed line.  Each row shows the results for different bin of redshift (as labeled).  The information inset in the panels give numbers of galaxies, $N$, along with the mean of the Gaussian fit(s), $\mu$ ($\mu_1$ and $\mu_2$ for the double-Gaussian fits), and $\Delta \mu = \mu_1 - \mu_2$, which shows the difference between the two means from the double--Gaussian fits (in units of dex).  The short, vertical lines show the means of the Gaussian fits, red lines show the fits to the higher-mass samples, and the blue lines show the fits to the lower-mass samples (the latter is shown in the both panels of each row for comparison). There is evidence for a bimodal distribution of galaxy s\sfrten\ for the higher-mass sample of galaxies at all redshifts in our sample.  Comparing the location of the means from the bimodal distribution for the higher-mass sample of galaxies with the peak from the unimodal distribution of the lower-mass sample of galaxies shows this appears to be the result of an increase in the number of galaxies with lower s\sfrten.  }
    \label{fig:kde_fits}
\end{figure}

We explore this further in Figure \ref{fig:kde_fits}, where we fit kernel-density estimators (KDEs) to the s\sfrten\ distributions in bins of redshift for the lower mass (left column, solid black lines) and higher mass galaxies (right column, solid black lines). We do not include galaxies at $9 < z < 12$ for the reason above that we lose the ability to accurately constrain \sfrten\ as the rest-frame optical nebular emissions shift beyond the wavelength coverage of NIRCam.  

The s\sfrten\ distributions for the lower mass galaxies in each bin of redshift appear consistent with unimodal distribution.  We therefore fit the s\sfrten\ distributions for the lower mass galaxies in each redshift bin  with a single Gaussian model (illustrated in each panel)  to the distributions for the lower mass galaxies in each redshift bin.  In each case, these are consistent with the KDEs. 

The s\sfrten\ distributions for the higher mass galaxies are more extended, and appear inconsistent with a unimodal distribution.  We fit a single Gaussian model to the each of distributions, and a Kolmogorov-Smirnov (KS) test fails to reject the hypothesis that these are drawn from the same parent distribution with an average of the $p$-values of each redshift bin  of $p = 0.60$. However, the distributions of s\sfrten\ for the higher mass galaxies appear bimodal. We therefore fit a double-Gaussian model to each of the s\sfrten\ distributions for the higher mass galaxies, illustrated in each panel of Figure~\ref{fig:kde_fits}.  In each panel of the right column, we list the means of the double Gaussian models, and we list the difference between the means $\Delta \mu = \mu_1 - \mu_2$.  Here we find that the KS test still does not reject the hypothesis that the distributions are drawn from the same parent distribution with an average of the $p$-values of each redshift bin of $p = 0.73$, a $\sim22\%$ increase in confidence over the single-Gaussian model. 

The fact that the distributions of s\sfrten\ for the higher mass galaxies appear bimodal means they need an additional component and this is not the case for the lower mass galaxies.  We can study this by comparing the location of the means (``peaks'') in the distributions of s\sfrten\ for the lower mass and higher mass samples as a function of redshift.  In each panel of Figure~\ref{fig:kde_fits}, we indicate the peak of the single component for the lower mass sample (left columns, blue vertical lines).  We also indicate the peaks of both the components for the higher mass sample (right columns, red vertical lines).  Comparing the peak values, we see that the peak of the s\sfrten\ distribution of the lower mass galaxies is always comparable -- i.e., within $\lesssim0.3$~dex -- to the peak of the higher-sSFR Gaussian for the higher mass galaxies. This means that the higher mass galaxies do not show evidence for higher s\sfrten\ compared to the lower mass galaxies.   Rather, the higher mass galaxies show a second component (modeled by a Gaussian here) with a peak at lower s\sfrten, shifted by 0.6~dex to 1~dex compared to the high-s\sfrten\ peak.   This means that the higher mass galaxies show evidence for a population of galaxies with lower s\sfrten.  
This appears to be driving the increase in the total scatter in \sfrten\ for galaxies at higher mass.  We therefore conclude that the increase in scatter is a result of there being more galaxies with lower s\sfrten\ (e.g., ``napping'' galaxies) rather than an increase in the relative number of galaxies with higher s\sfrten\ (e.g., there is not an excess of ``bursting'' galaxies).  We discuss the implications of this below. 

\section{Discussion}\label{section:discussion}

The results presented here favor a scenario where the mode of star formation in galaxies in the early Universe is more variable than that observed in galaxies at later cosmic times.   
The variable nature of star formation in early galaxies has been predicted in some simulations \citep[e.g.,][]{sun:23,wilkins:23b,yung:23b}, and has been proposed as one of the solutions to account for the higher number density of high redshift galaxies at the bright end the UV luminosity function  \citep[e.g.,][]{mason:23, finkelstein:22b, chworowsky:23a}.  Other studies based on SED modeling and measurements of galaxy EW distributions also favor ``burstier'' star-formation in high redshift galaxies \citep{endsley:22,endsley:23a,ciesla:23}.  This runs counter to studies at relatively lower redshifts that favor more slowly-evolving star-formation histories \citep[e.g.,][]{papovich:11,reddy:12}.  The fact that we find galaxy SFRs have more scatter on short timescales, $\lesssim$100~Myr, implies that the mechanism driving this increased variability also acts on these shorter timescales.  However, ``burstier'' may not be an accurate description for the SFHs of these galaxies.  The galaxies do not have excessively high s\sfrten\ values, but rather show an increase in the frequency of galaxies with low s\sfrten\ values.  Therefore the higher variability in galaxies in the EoR is driven by one or more physical mechanisms that interrupt otherwise ``normal'' star formation.      Here, we discuss the implications on galaxy growth at $z > 6$ (in the EoR) as a result of the increased variability in the galaxy SFRs.  We also discuss how our results compare to theoretical predictions and to the findings from other recent studies of galaxies at these epochs.   

\subsection{An Increase in the Scatter in the SFRs of Galaxies at the Epoch of Reionization }

We see an increase in intrinsic scatter ($\sigma_\mathrm{int}$) along the SFMS measured with SFRs on 10~Myr timescales at increasing redshift, and we see that the scatter in \sfrten\ (\sigmaten) also increases with increasing stellar mass. This result is similar to the findings from many observational studies of galaxies at these epochs.  One finding is that star formation must become increasingly more variable as we look to higher redshifts, based on the strength of galaxy nebular emission lines. \citet{endsley:23a} find that emission line equivalent width (EW) distributions are broader, ranging to larger values, for galaxies at $6<z<9$ in the JADES field \citep{rieke:23, williams:23}. An explanation for this is that the galaxies contain a higher relative number of OB stars. \citet{endsley:23a} also find that a considerable subset of the SEDs in the CEERS sample show strong evidence for Balmer/4000\AA\ breaks, which indicates there is a substantial fraction of galaxies observed in CEERS have SEDs dominated by older ($\gtrsim 100$~Myr) stellar populations. \citet{endsley:23a} concluded that these observations favor an increase in the variability of star formation at this epoch. This agrees with our findings.  Moreover, \citet{endsley:23a} provide an independent analysis as they fit the galaxy SEDs using \textsc{BEAGLE} with a constant SFH \citep{chev_char:16}.  Therefore this is further support for an increase in the scatter of SFRs of galaxies in the EoR.

Studies of the UV luminosity function (UVLF) and its evolution  at $z\gtrsim7$ have invoked a high variability in star formation to explain the excess number density of galaxies at the bright end of the UVLF \citep[e.g.,][]{finkelstein:23, mason:23}. \citet{shen:23} conclude that an increase in the scatter of UV absolute magnitudes would account for the differences of the UVLF, where they require of $\sigma_{M_\mathrm{UV}} \approx 0.75$~mag at the bright end and $\sigma_{M_\mathrm{UV}} \approx 1.5$~mag at the faint end of the UVLF.   In the JADES fields, \citet{ciesla:23} measure $\sigma_{M_{UV}} \approx 1.2$ mag at $z>9$ for galaxies with $-16.6 > M_\mathrm{UV} > -18.4$, consistent with this idea.  Comparing to our work here, we calculate a scatter in the SFR for galaxies at $7<z<9$ of $\sigmaten \approx 0.6$ dex for high--mass galaxies.   Assuming $\sigma(\log \mathrm{SFR}) \sim 0.4 \cdot \sigma(M_\mathrm{UV})$, this corresponds to $\sigma_{M_{UV}}$ $\sim 1.5$ mag.   Therefore, the scatter we derive appears sufficient to account for the higher number density of objects at the bright end of the UVLF. We defer a detailed comparison of how the increase in the scatter of the SFRs impacts the UVLF and the stellar mass function (SMF)  to a future study. 

At the high-mass end of the SFMS, we observe a bimodality in the distribution of s\sfrten\ (Figure~\ref{fig:kde_fits}) implying that galaxies with log($M_\ast/M_\odot$) $\gtrsim 9.3$ are driving the intrinsic scatter along the SFMS on 10 Myr timescales.   Furthermore, our results show this is apparently a result of an increase in the frequency of galaxies with low sSFRs.   There is growing evidence from other studies that support this conclusion.  \citet{ciesla:23} presented evidence for two populations of star-forming galaxies at $7<z<9$ based on the SFR ``gradient''.  One population has a SFR gradient moving them along the SFMS, while the other population has a gradient moving them below the SFMS (toward lower sSFRs).   Qualitatively, the first population is consistent with the galaxies we show in Figure \ref{fig:sfms_SED} that have SFRs that place them on the SFMS. The second population includes the kinds of galaxies we show in Figure \ref{fig:lowsfr}, where we are observing them during a recent increase in star formation, or in a declining phase (so-called ``napping'' galaxies, which have experienced a cessation in star formation).   The latter population includes galaxies similar to those seen in recent studies.  For example, \citet{looser:23a} report on a galaxy, JADES-GS-z7-01-QU, at $z=7.3$ with no nebular emission lines, which they conclude ceased forming stars 10-20 Myr earlier after a short, substantive episode of star formation.  \citet{looser:23b} present additional evidence for galaxies at $5 < z < 11$ in this ``napping'' stage (what they call ``mini-quenched'' galaxies).   \citet{gelli:23} argue that $\sim 30\%$ of galaxies with stellar masses $\log M_\ast/M_\odot \lesssim 9.5$ at these redshifts have SEDs indicating they are in this ``napping'' mode, experiencing lower sSFRs.    There is therefore growing evidence that galaxies in the EoR experience active periods of star formation, followed by periods of cessation. 

\begin{figure*}
    \centering
    \includegraphics[width=0.8\linewidth]{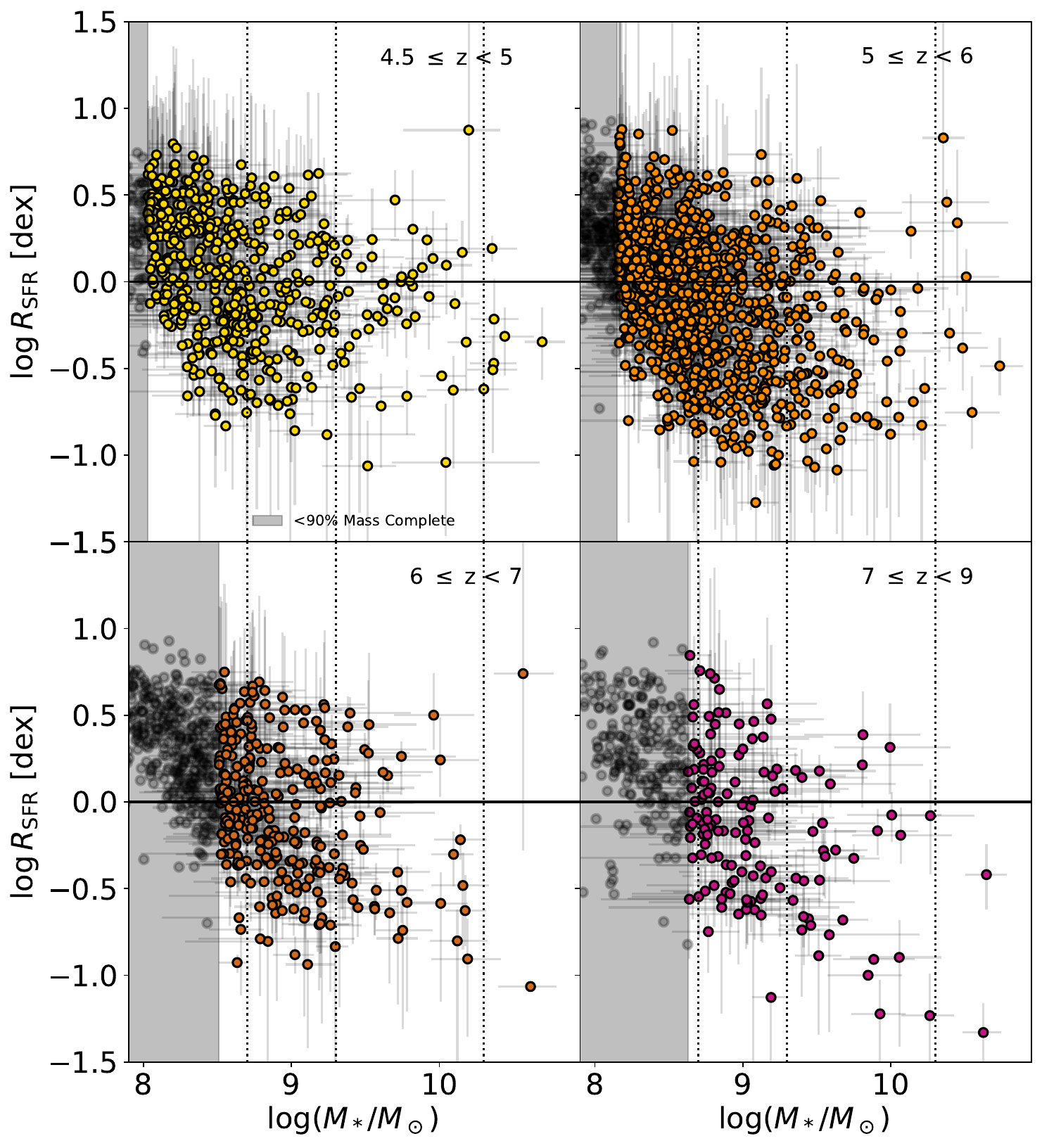}
    \caption{Ratio of the SFRs averaged over 10~Myr (\sfrten) to the SFRs averaged over 100~Myr (\sfrcen) as a function of stellar mass and redshift, $\log R_\mathrm{SFR}$ = $\log \sfrten - \log \sfrcen$.    Each panel shows galaxies in a different redshift bin (as labeled; colored as in Figures \ref{fig:sfrm_split} and \ref{fig:sfrcomp}.  We divided the sample into a lower mass (log($M_*/M_\odot$) $\in [8.7,9.3]$) and a higher mass sample (log($M_*/M_\odot$) $\in [9.3,10.3]$), delineated by the dotted, vertical lines. The thick horizontal solid line shows where the SFRs are equal, \sfrten\ = \sfrcen.\  The gray shaded region in each panel shows where our sample drops below 90\% completeness in stellar mass. 
    }
    \label{fig:sfrdiff}
\end{figure*}

\begin{figure}
    \centering
    \includegraphics[width=\linewidth]{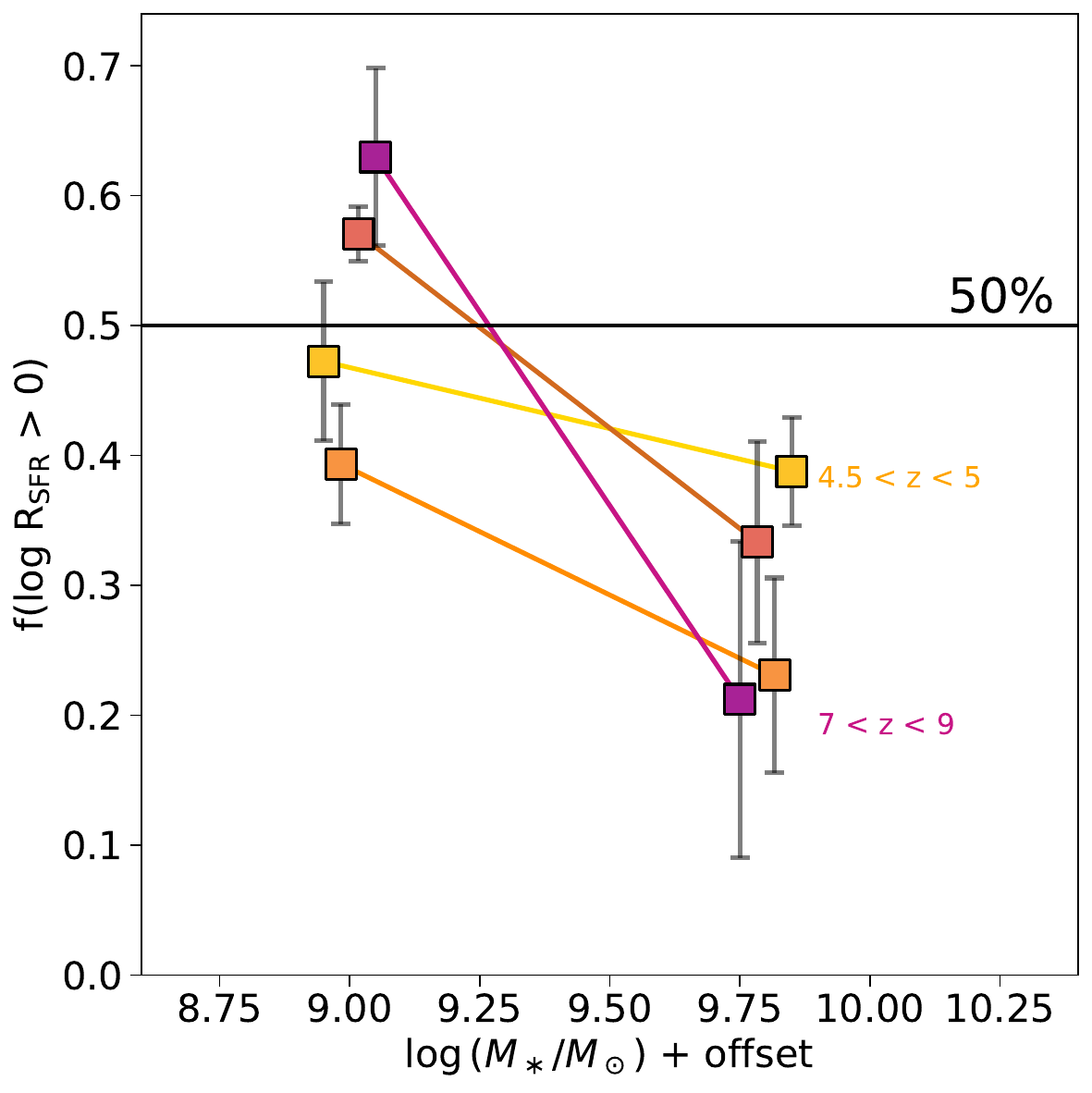}
    \caption{The fraction of galaxies that have ratios $R_\mathrm{SFR} \equiv \sfrten/\sfrcen$ greater than unity, $f(\log R_\mathrm{SFR} > 0)$, which is an estimate of the ``duty cycle'' of star formation in galaxies. The markers show galaxies in different redshift bins, colored as in Figure \ref{fig:sfrdiff} going from $4 < z < 5$ (yellow) to $7 < z < 9$ (purple), which are labeled for clarity. The errors shown are estimated using a bootstrapping method in which we resample each galaxy (at $4.5 \leq z \leq 9$) and remeasure the fractions to account for uncertainties in redshift, stellar mass and SFR. At all redshifts, the higher-mass samples have a higher fraction of galaxies with $\log R_\mathrm{SFR} < 0$, but the change from the lower-mass to higher-mass galaxies is the most dramatic for galaxies at higher redshifts.  }
    \label{fig:sfrdiff_total}
\end{figure}

\subsection{The Duty Cycle of Star Formation in the Epoch of Reionization}\label{section:DC}

The scatter in the SFMS implies that galaxies spend some period of time in starburst phases \textit{and} in ``napping'' phases.  The fraction of the galaxy population in this phase places an estimate on this time.  We then define a ``duty cycle'' of star formation as the relative amount of time galaxies spent in the star-forming phase (when they lie on the SFMS).  

We estimate the fraction (of galaxies in active star-forming stages) using the ratio of the SFRs averaged over 10 Myr compared to the SFRs averaged over 100 Myr.  We define this ratio as $R_\mathrm{SFR} \equiv \sfrten/\sfrcen$, similar to the quantity studied by \citet{looser:23b}.   We then use this ratio to quantitatively identify galaxies in a declining star-forming stage, with $\log R_\mathrm{SFR} < 0$, and galaxies that are in active star-forming stages, with  $\log R_\mathrm{SFR} \geq 0$.  Figure \ref{fig:sfrdiff} shows the ratios of $\log R_\mathrm{SFR}$ as a function of stellar mass for galaxies in the different redshift bins.  Figure \ref{fig:sfrdiff} illustrates that, lower mass galaxies include galaxies with high and low values of $\log R_\mathrm{SFR}$ in nearly equal amounts, with a distribution centered near unity.   In contrast, higher mass galaxies appear to have a higher fraction of objects with low ratios, $\log R_\mathrm{SFR} < 0$.  Quantifying this trend, at redshifts $4.5<z<9$, $>60\%$ of higher mass galaxies with $\log M_\ast/M_\odot > 9.3$ have low ratios, $\log R_\mathrm{SFR} < 0$.   That is, at these redshifts and stellar masses, galaxies are more likely to be in a phase of lower SFR than in an active star-forming phase.    Generally this implies that \textit{on average} lower mass galaxies have SFHs that are constant on 10-100~Myr timescales while higher mass galaxies have SFHs that have declined over these timescales. 

Figure \ref{fig:sfrdiff_total} illustrates these trends by showing the fraction of galaxies with high $R_\mathrm{SFR}$, that is, $f(\log R_\mathrm{SFR} > 0)$ (top panel) in bins of stellar mass and redshift. Lower-mass galaxies with $\log{(M_\ast/M_\odot)} < 9.3$, have near equal fractions above and below the unity line, with $f \approx 0.5$.  It is interesting to see if these trends continue or change at lower masses, where there is some evidence that the fraction of galaxies with $\log R_\mathrm{SFR} > 0$ increases, indicating a higher fraction of bursts \citep[e.g.,][]{looser:23b}.   Higher mass galaxies with $\log{(M_\ast/M_\odot)} > 9.3$ have a lower fraction of galaxies with high $R_\mathrm{SFR}$ values with $f(\log R_\mathrm{SFR} \geq 0)$ of $<40\%$.  This fraction decreases with increasing redshift, falling to $\simeq 15-30\%$ at $7<z<9$. 

We take the fraction of galaxies above the unity relation, $\log R_\mathrm{SFR} > 0$, which yields an estimate of duty cycle in a redshift and mass range.  We further can estimate the average amount of time galaxies spend in the active star-forming stage by multiplying the duty cycle with the time period ($\Delta t$) spanned by the redshift range in each bin,
\begin{equation}\label{eq:dc_bursty}
    t(\log R >0) = f(\log R >0) \times \Delta t.
\end{equation}
 We list the duty cycles and the time period of active star formation for lower mass and higher mass galaxies in bins of redshift over the range $4.5 < z < 9$ in Table~\ref{table:dutycycle}. 

\begin{deluxetable*}{c@{~~~~~~~~~~}c@{~~~~~~~~~~}ccc@{~~~~~~~~~}ccc}
\caption{Duty cycle of star formation for galaxies }\label{table:dutycycle}
\tablehead{\colhead{} & \colhead{} & \multicolumn{6}{c}{Galaxies with stellar mass} \\ \colhead{} & \colhead{} & \multicolumn{3}{c}{$9.3 \leq \log M_\ast/M_\odot \leq10.3$~~~~~~~~~~} & \multicolumn{3}{c}{$8.7 \leq \log M_\ast/M_\odot < 9.3$} \\ \colhead{Redshift range~~~~~~~~~~} & \multicolumn{1}{c}{$\Delta$t [Myr]~~~~~~~~~~~~} & \colhead{N} & \colhead{~~~$f(\log R_\mathrm{SFR}>0)$} & \colhead{$t = f\times \Delta t$~[Myr]~~~~~~~~~~} & \colhead{N} & \colhead{$f(\log R_\mathrm{SFR} > 0)$} & \colhead{$t = f \times \Delta t$~[Myr]}
}
\startdata
    $4.5 < z \leq 5$ & 172 & 19 & $39\pm 4$\% & $67\pm 7$ & 137 & $47\pm 6$\% & $81\pm 11$ \\
    $5 < z \leq 6$ & 273 & 27 & $23\pm 8$\% & $61\pm 21$ & 262 & $39\pm 5$\% & $107\pm 13$ \\
    $6 < z \leq 7$ & 210 & 14 & $33\pm 8$\% & $70\pm 16$ & 122 & $57\pm 2$\% & $120\pm 5$ \\
    $7 < z \leq 9$ & 307 & 7 & $21\pm 13$\% & $45\pm 27$ & 51 & $63\pm 7$\% & $132\pm 15$ \\
\enddata
\tablecomments{$\Delta$t is the cosmic time spanned by each redshift range.  $f$ is the duty cycle when galaxies are in the ``star-forming'' mode and $t(\log R >0)$ is the average duration of time galaxies spend in active periods of star formation stars. We show all timescales in Myr.}
\end{deluxetable*}

Regarding the duty--cycle measurements we see several interesting trends. First, higher mass galaxies spend a shorter period in an active star-forming mode than lower mass galaxies.  This is more significant at higher redshifts, $z > 5$ where the time spent in active stages for the higher mass galaxies lies more $>2\sigma$ away from the value for the lower mass galaxies.  The active star-forming stages for higher mass galaxies have an average duration of  $\simeq 50-70$~Myrs.  Lower mass galaxies spend $\simeq 80-130$~Myrs in this active mode.   

These timescales in the active star-forming stage for high redshift galaxies are similar to the those predicted from recent predictions  for (so-called) ``feedback-free starbursts'' \citep[FFBs, see ][]{dekel:23, li:23}.  FFBs are predicted to occur when the gas densities of galaxies lead to sufficiently short free-fall times such that galaxies can form bursts of stars on timescales shorter than the lifetime of massive O-stars.  This FFB model also predicts that galaxies at $z>10$  experience episodes of star formation that span $\sim 80$~Myr timescales. \citet{dekel:23} demonstrate that feedback from supernova and stellar winds typically does not begin affecting star formation in neighboring gas clouds for $\sim2-3$~Myrs after the initial burst, and subsequent bursts are fed by continuous gas accretion over a $\sim80$~Myr timescale \citep{dekel:23}. Our measurement of the duty cycle suggests that if the FFB mechanism is the limiting physical mechanism for star formation at these epochs, then it should continue throughout the EoR, at least down to $z\sim6$. One possible way to test these predictions is to measure accurate gas densities and gas fractions of galaxies at these redshifts. 

The duty cycle for active star formation, $\log R_\mathrm{SFR} >0$, estimated in this study appears to be  independent of redshift over the range $4.5 < z < 9$. The average time spent in a period of active star formation is $\simeq 65$~Myr for higher mass galaxies and $\simeq 110$~Myr for lower mass galaxies.   As noted above, this is similar to the predictions for the FFB scenario (though alternative theories exist, see Section~\ref{section:discussion:models} below).  The FFB scenario also predicts that galaxies should spend time in a ``lull'' between active star-forming stages, where we predict these lulls should last from $\sim 100$ to 250 Myr based on the duty-cycle values we derive (see Table~\ref{table:dutycycle}).  

One important finding from our work is that the active periods of star formation are shorter and these lull periods are longer for the higher mass galaxies ($9.3 \leq \log M_\ast/M_\odot \leq 10.3$) than lower mass galaxies.   However, we also find that these lull periods become \textit{shorter} with decreasing redshift as the duty cycles increase (see Table~\ref{table:dutycycle}).  This naturally leads to more smooth star-formation if the time period of the ``lulls'' continue to shorten as cosmic time increases (i.e., moving toward lower redshifts)  such that the duty cycle trends toward unity.  This accounts for  more stochastic star formation at high redshifts, followed by more smooth star formation at lower redshifts.   This is similar to the conclusion of \citet{ciesla:23}, but we find the transition from bursts to smooth star-formation is more gradual and continues throughout the redshift range here, $4.5 < z < 9$.   Any physical mechanism or mechanisms regarding gas accretion, star-formation, and feedback in these galaxies must account for these observations. 

%
\subsection{Physical Implications from Simulations}\label{section:discussion:models}
Studies of cosmological models and simulations have identify multiple pathways that fuel and regulate star formation during and after the Epoch of Reionization.  The main physical picture involves the growth of dark-matter halos, gas accretion and cooling, star formation, and feedback.  Current theoretical studies have yet to reach convergence on the physical processes and the strength of various effects.   Regarding feedback, many studies rely on AGN, which can  effectively disrupt gas accretion and drive cold gas from galaxies, causing a cessation of star formation \citep[e.g., ][]{hopkins:10}.   However, \citet{wilkins:23a} demonstrated that AGN feedback is less effective in reproducing observations toward the end of the EoR ($4.5<z<6$).  This is supported by observational studies \citep[e.g., ][]{smit:16} that demonstrate the need for rising SFHs and AGN at $z\lesssim5$ to reproduce observations. However, at higher redshifts, these mechanisms are unable to reproduce the abundance of galaxies observed to have recently shut down their star formation \citep[see, e.g., ][]{heckman_best:14, wilkins:23a}.  Therefore AGN may be insufficient to regulate star formation in these galaxies. Moreover, to explain the scatter in the SFRs we find, any AGN feedback would need to be episodic on timescales of $\sim$100-200~Myr.

We can also use predictions from simulations and SAMs to understand the global impact of physical effects.   Figure~\ref{fig:sfrcomp} compares predictions from the SC SAM model \citep{yung:23b} and from the FLARES simulation \citep{wilkins:23b}.  Both the SC SAM and FLARES simulation include prescriptions for gas accretion and cooling, star formation and stellar feedback.   These all contribute to the scatter of galaxy SFRs on the SFMS.   These appear to recover the scatter we measure in the SFRs on 100~Myr timescales (\sigmacen).  As illustrated in Figure~\ref{fig:sfrcomp} we measure $\sigmacen \approx 0.2-0.3$~dex, which agrees with both the SC-SAM and FLARES predictions for the scater on these timescales   However, both the SC-SAM and FLARES predictions fall short of the scatter we measure on 10 Myr timescales, where we measure $\sigmaten \approx 0.4-0.6$~dex while the predictions for the scatter on 10~Myr timescales is very similar to the predictions for this on 100~Myr timescales. This suggests that the timescales of physical effects pertaining to galaxy star-formation is too long in these models and needs to vary more quickly to account for our observations.  One possibility is the effect of FFBs which are predicted to act on shorter timescales under the physical conditions expected for these high redshift galaxies (\citealt{dekel:23} and \citealt{li:23}, and discussion in Section~\ref{section:DC}). 

Other studies have demonstrated that multi-channel stellar feedback processes are sufficient in regulating star formation and impact the specific SFRs of galaxies.  For example, \citep{mason:23} demonstrate these processes are able to reproduce the evolution of the UVLF at high redshifts (see also \citealt{shen:23}). Additionally, \citet{wilkins:23a} find that the effects of stellar feedback causes more massive star-forming galaxies at these redshifts tend to have lower sSFRs than lower mass galaxies.  These predictions are consistent with  our findings (Figure~\ref{fig:ssfr_z}).  
\section{Conclusions}\label{section:etfin}
In this paper, we study the relation of galaxy stellar mass and SFR at high redshifts.  We focus on trends at $4.5 < z < 9$ and extending this in some cases to $9 < z < 12$.  We study the SEDs of $>$1800 galaxies using  multi-wavelength data from \JWST/NIRCam from CEERS, combined with legacy \HST/ACS and WFC3 imaging from CANDELS.  We model the SEDs with stellar--population models that span a range of parameters, and include nebular emission, and flexible SFHs.   We show that the multiwavelength dataset is sensitive to changes in the nebular emission lines of galaxies which tracks the SFRs on timescales of 10~Myr (the lifetime of O-type stars) and to changes to the relative amount of rest-UV emission that tracks the SFRs on timescales of 100~Myr.  This allows us to compare the SFRs on short, 10~Myr timescales compared to longer, 100~Myr timescales.   

We then use these results to measure and to analyze the evolution of the SFMS out to the redshifts.  We implement an MCMC routine utilizing a two-part likelihood function that simultaneously measures the slope, normalization and intrinsic scatter of the SFMS as a function of redshift.   We study the distribution of galaxies in SFRs using \sfrten\ and \sfrcen, and in the s\sfrten\ spaces.  We use the ratio of $R_\mathrm{SFR} = \sfrten/\sfrcen$ to estimate evolution of galaxy SFHs and to estimate a duty cycle of star formation, all as functions of redshift and stellar mass.  A summary of our primary findings are as follows.

%
        
The \jwst/NIRCam multiwavelength photometry is well-suited for measuring SFRs over 10 Myr timescales. This is primarily true for the sample of galaxies with $z\lesssim9$ where the photometry covers both the rest-UV and Balmer regions of the observed SEDs, and samples the strength of emission lines in the rest-optical (\oiii, \hb, etc).   This allows better constraints on stellar mass, the SFR, and the SFH in galaxies.  Figures~\ref{fig:highsfr}--\ref{fig:lowsfr} show examples of these fits and how the SFH constraints depend on the features of the model fits to the multiwavelength SEDs.  
    %
       
There is minimal evidence that the evolution in the normalization of the SFMS changes from $4.5 < z < 12$ when using SFRs averaged over 10 Myr or 100 Myr timescales (Figure~\ref{fig:sfrm_full}). 
    %
    %
       
The scatter of the SFMS is larger when using SFRs averaged on 10~Myr timescales (\sigmaten) compared to the SFRs averaged on 100~Myr timescales (\sigmacen).   We find $\sigmaten \approx 0.3-0.5$~dex compared to $\sigmacen \approx 0.1-0.2$~dex at with no significant evolution in redshift for $4.5 < z < 9$ (Figures~\ref{fig:sfrm_split_100} and \ref{fig:sfrm_split}).   The values of \sigmacen\ do not appear to depend on stellar mass.   However, while the scatter in the SFRs on 10~Myr timescales is also apparently constant as a function of stellar mass from $4.5 < z < 6$, it shows evidence that it increases with increasing stellar mass at $6 < z < 9$ (Figure~\ref{fig:sfrcomp}).
    %
        
The distributions of specific \sfrten\ in galaxies evolve with both mass and redshift, but the s\sfrten\ distributions for higher mass galaxies are lower than those for lower mass galaxies by $\approx$0.3~dex at all redshifts, $4.5 < z < 9$ (Figure~\ref{fig:ssfr_z}). 
    %
        
The distribution of log(s\sfrten) for lower mass galaxies, $8.7 \leq \log M_\ast/M_\odot \leq 9.3$, is consistent with a single Gaussian with a mean log(sSFR) that increases slightly with redshift from $\log(s\sfrten)/\mathrm{yr}^{-1}) = -8.4$ at $z\approx 5$ to $-8.0$ at $z\approx 8$.   Higher mass galaxies, $9.3 < \log M_\ast/M_\odot \leq 10.3$ have a larger scatter in their s\sfrten.  This is because their log sSFR distributions are apparently bimodal, which we model as  a double Gaussian.  Interestingly the double Gaussian models have a high sSFR peak (mean) that is consistent with that of the (single) Gaussian models of lower mass galaxies, while the second Gaussian model for the higher mass galaxies is shifted to lower sSFRs (Figure~\ref{fig:kde_fits}).  Therefore the increase in the SFR scatter is not because of a excess of ``bursting'' galaxies, but rather because of an increase in the fraction of galaxies experiencing a lull in their SFRs at these redshifts.  
    %
    
We use the ratio of \sfrten--to--\sfrcen\ to describe trends in galaxy SFHs.   We define $\log R_\mathrm{SFR} = \log \sfrten - \log \sfrcen$ as a ratio of the current SFR (on 10~Myr timescales) to the past average (on 100~Myr SFR).  The distribution of $R_\mathrm{SFR}$ is centered near unity for lower mass galaxies, while for higher mass galaxies the distribution shifts to $\log R_\mathrm{SFR} < 0$  (Figure~\ref{fig:sfrdiff}).  Generally this implies that \textit{on average} lower mass galaxies have SFHs that are constant on 10-100~Myr timescales while higher mass galaxies have SFHs that have declined over these timescales. 
    %
    
    We measure a duty cycle for the phase of active star formation in galaxies as the fraction of galaxies that have $\log R_\mathrm{SFR} > 0$ as a function of stellar mass and redshift.   For higher stellar mass galaxies, $9.3 \leq < \log M_\ast/M_\odot \leq 10.3$, the duty cycle is $\approx$30\%, and increases from 10--30\% at $7< z < 9$ to $35-43$\% at $4.5 < z < 5$.   Lower mass galaxies, $8.7 \leq \log(M_\ast/M_\odot) < 9.3$, the duty cycles are higher, averaging 40--60\% with little change in redshift (Figure~\ref{fig:sfrdiff_total} and Table~\ref{table:dutycycle}).
    %
    
    Multiplying the time spanned by the redshift range and the duty cycle leads to an estimate for the duration of time galaxies spend in active star-forming stages  (Table~\ref{table:dutycycle}).  For higher stellar mass galaxies, $9.3 \leq < \log M_\ast/M_\odot \leq 10.3$, this period of active star formation averages $\approx$60~Myr with no evidence for evolution over $4.5 < z < 9$.   For lower mass galaxies the periods of active star formation are longer, averaging $\approx$110~Myr over this redshift range.   At least for the higher mass galaxies, the fact that the timescales of active star formation are roughly constant, but the duty cycle of star formation increases implies the periods of \textit{inactivity}, when galaxies experience a lull in star formation, must decrease with cosmic time (decreasing redshift).   This naturally implies that galaxies transition from a period of higher variability in their star formation at higher redshift to more smooth star formation at lower redshifts when the inactive, lull periods become short.  We argue we are witnessing this transition from $z\sim 9$ to $z \sim 4.5$.    
%

Future studies of the star formation histories and properties of high redshift galaxies with \jwst\ and other facilities will test some of the findings in this work.  Specifically, while current theoretical predictions account for some of our findings (Section~\ref{section:discussion}), they are unable to account for the higher variability in star formation on shorter, 10~Myr timescales.   To explain these findings requires new physics, possibly with novel ideas such as FFBs.  These theories make predictions for the gas fractions and gas densities of galaxies and their dependence on mass, SFR and redshift.  All of these can be tested with future observations.    

\begin{acknowledgements}

We wish to thank all our colleagues in the CEERS collaboration for their hard work and valuable contributions on this project.  We also thank all the scientists, engineers, and administrators that made \jwst\ the incredible time machine that it is. JWC acknowledges support from the NASA Headquarters under the Future Investigators in NASA Earth and Space Science and Technology (FINESST) award 22-ASTRO22-0264. CP thanks Marsha and Ralph Schilling for generous support of this research.   This material is based in part upon work supported by the National Science Foundation under grants AST-2009632. Portions of this research were conducted with the advanced computing resources provided by Texas A\&M High Performance Research Computing (HPRC, \url{http://hprc.tamu.edu}).  This work benefited from support from the George P. and Cynthia Woods Mitchell Institute for Fundamental Physics and Astronomy at Texas A\&M University.    This work acknowledges support from the NASA/ESA/CSA James Webb Space Telescope through the Space Telescope Science Institute, which is operated by the Association of Universities for Research in Astronomy, Incorporated, under NASA contract NAS5-03127. Support for program No. JWST-ERS01345 was provided through a grant from the STScI under NASA contract NAS5-03127.

\end{acknowledgements}
\appendix
\section{Impact of SFR--Stellar Mass Covariance on the Scatter of the SFMS}
\label{section:appendix}

The covariance between the SFR and $M_*$ for typical galaxies in our sample are correlated with a positive covariance (see Figure~\ref{fig:sfrm_full} and Section~\ref{section:sfms}).   For star-forming galaxies this covariance arises from the fact that the redder ``color'' of a galaxy is correlated with both quantities.  That is, a redder color corresponds to an increase in the stellar-mass--to--light ratio (and thus increases the stellar mass) and a redder color typically indicates higher levels of dust attenuation which increases the SFR.   We  therefore test how this covariance affects the measurement of scatter along the SFMS.  First, we test the fidelity of our two-function likelihood method described in Section \ref{section:sfms} by applying the methodology with scatter only in the dependent variable, i.e., the SFR.   Second, we test how well we are able to recover the scatter when including the effects of the covariance between the dependent and independent variables, i.e., SFR and stellar mass.  

We use a model where galaxies have a linear relation between the variables such that 
\begin{equation}
    y_i = m x_i + b + \epsilon_i, 
    \label{eq:a1}
\end{equation}
where the index $i = \{1..N\}$ represents the number of data points (e.g., $N$ galaxies), and $x_i$ is the independent variable and $y_i$ is the dependent variable. Here, $m$ and $b$ are the slope and normalization of the relation for the sample, and $\epsilon_i \sim$ N(0, $\sigma^2$) is the amount of scatter experienced by datum $i$ given a sample scatter of $\sigma$. We then run a MCMC, implemented with \textsc{emcee} \citep{foreman-mackey:13}, to infer $\hat{\theta} = $ \{$\hat{m}, \hat{b}, \hat{\sigma}$\}, the slope, normalization and intrinsic scatter, respectively. \textit{NB: we adopt the notation here where the ``hat" on variables, i.e., $\hat{y}$, distinguishes variables that are estimates compared to true quantities. The symbol ``$\sim$'' means ``is distributed as'', i.e., $y_i^\prime \sim N(0,\hat{\sigma}^2)$ reads ``$y_i^\prime$ is distributed as a normal distribution''.}

Here, we want to separate the slope and normalization from the intrinsic scatter in order to accurately describe the \textit{absolute} intrinsic scatter, or the intrinsic scatter that reproduces the distribution $|N(0,\sigma^2)|$, while simultaneously inferring the full posterior space for $\hat{\theta}$. Because we are assuming that the variables are distributed along a linear relation and the scatter about that relation is distributed normally, this requires both a linear likelihood function and a $\chi^2$-likelihood function to accurately describe the distribution of the data \textit{about} the linear relation. This is easily achieved by separating the two distributions and calculating their likelihoods separately. 

Therefore, for each iteration of the MCMC fit, we split Equation \ref{eq:a1} to first derive the slope and normalization such that:
\begin{equation}
    \Vec{\mathbf{y}} = \hat{m} \Vec{\mathbf{x}} + \hat{b} \label{eq:a2}
\end{equation}
where $\Vec{\mathbf{x}}$ and $\Vec{\mathbf{y}}$ in our case contains the mass and SFR values for the full set of galaxies.  We then calculate $\hat{\Vec{\mathbf{y}}} = \hat{m} \Vec{\mathbf{x}} + \hat{b}$, which is the estimate of the SFRs given the parameters $\hat{m}$ and $\hat{b}$. In the second step we calculate
\begin{equation}\label{eq:a4}
    \Vec{\mathbf{y}} - \hat{\Vec{\mathbf{y}}} = \Vec{\mathbf{y}^\prime} \sim N(0,\hat{\sigma}^2).
\end{equation}

This is important as the measured estimates are used to  infer the scatter in Equation \ref{eq:a4}. Equation \ref{eq:a4} then describes the estimated measurement of the intrinsic scatter ($\hat\sigma$) along the linear relation. Subtracting off the linear relation leaves only points scattered about $y~=~0$ and we model as a normal distribution. 

The likelihood equations for these two functions then become:
\begin{equation}\label{eq:linlike}
    \mathcal{L}_1(\hat{m},\hat{b}|\Vec{\mathbf{x}}) = \prod_{i=1}^N \hat{m} x_i + \hat{b},
\end{equation}
\begin{equation}\label{eq:chilike}
    \mathcal{L}_2(\hat{\sigma} | \Vec{\mathbf{y}^\prime}) = \prod_{i=1}^N \frac{(y_i^\prime - \epsilon_i)^2}{\hat{\sigma}^2} = \frac{1}{\hat{\sigma}^{2N}} \prod_{i=1}^{N} (y_{i}^\prime - \epsilon_i)^2,
\end{equation}
where $\epsilon_{i} \sim N(0,\hat{\sigma}^2)$ and we assume a chi-squared--likelihood under the assumption that the scatter about the linear function follows a normal distribution. Taking the natural logarithm ($ \ln{\mathcal{L}} =  \ell$) of the two functions provides the log-likelihood function used in our MCMC routine. They have the form:
\begin{equation} \label{eq:loglin}
    \ell_1(\hat{m}, \hat{b} | \Vec{\mathbf{x}}) = \ln\left[ \prod_{i=1}^N \hat{m} x_i + \hat{b} \right] = \sum_{i=1}^N \ln{\big( \hat{m}x_i ~+~ \hat{b} \big)},
\end{equation}
\begin{equation} \label{eq:logchi}
    \ell_2(\hat{\sigma} | \Vec{\mathbf{y}^\prime}) ~=~ \ln\left[\frac{1}{\hat{\sigma}^{2N}} \prod_{i=1}^{N} (y^\prime_{i} - \epsilon_i)^2\right]~=~ -2N \ln{\hat{\sigma}} + \sum_{i=1}^{N} \ln{\left[(y^\prime_{i} - \epsilon_i)^2\right]},
\end{equation}
The final form of the likelihood function must then be the sum of Equations \ref{eq:loglin} and \ref{eq:logchi},
\begin{equation} \label{eq:loglike}
    \ell(\hat{\theta} | \Vec{\mathbf{x}}, \Vec{\mathbf{y}^\prime}) ~=~ \ell_1\ + \ell_2\ ~=~ \sum_{i=1}^N \ln{\left( \hat{m}x_i ~+~ \hat{b} \right)}\ -2N \ln{\hat{\sigma}} + \sum_{i=1}^{N} \ln{\left[(y^\prime_i - \epsilon_i)^2\right]},
\end{equation}
and is the total likelihood function that we implement in our MCMC routine. 

\begin{figure*}
    \centering
    \includegraphics[width=\linewidth]{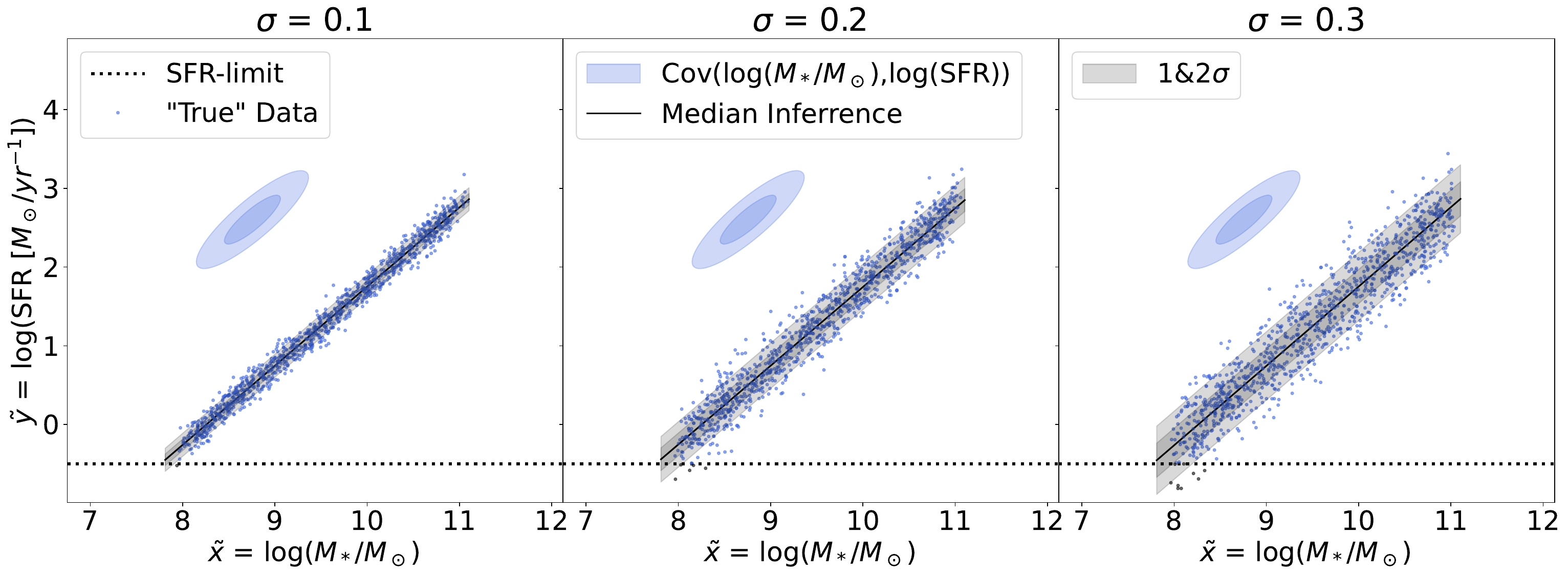}
    \caption{Simulated SFR-M$_\ast$ relations assuming different values of the intrinsic scatter. Blue circles show the simulated data and the solid black line shows the median best-fit from the MCMC routine, with 1 and 2 $\sigma$ shown as the gray-shaded regions. The blue ellipses show the 1 and 2 $\sigma$ covariance matrix from which the simulated data is drawn.}
    \label{fig:app_fits}
\end{figure*}

\begin{figure*}
    \centering
    \includegraphics[width=\linewidth]{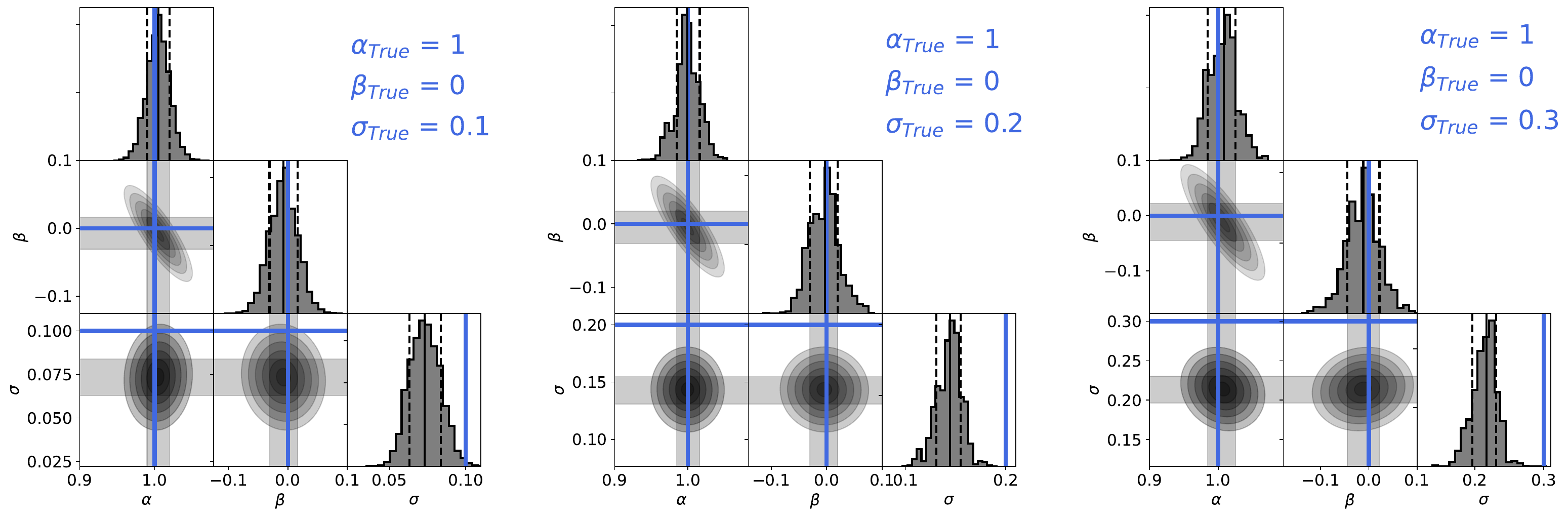}
    \caption{Corner plots from the MCMC routine showing the relationships between the posteriors of the slope ($\alpha$), normalization ($\beta$) and the intrinsic scatter ($\sigma$). From left to right, we show the results assuming $\sigma=0.1$, $\sigma=0.2$ and $\sigma=0.3$.}
    \label{fig:app_corner}
\end{figure*}

To test the ability of this routine to measure the SFR-M$_\ast$ relation, we start by generating $N=$20,000 random data points for $\Vec{x}$ from a uniform distribution with $8\leq x \leq 11$.   We then compute $y_i = \alpha (x_i - M_0) + \beta $, where $\alpha~=~1$, $\beta~=~0$, and $M_0=8.25$ to allow the relation to span similar values as our data.  This provides the set of {$x_i, y_i$} for the $i=1\ldots N$ points.  To introduce uncertainties on ``stellar mass'' and ``SFR,'' we generate a correlation matrix using the $P(M_\ast/M_\odot,\mathrm{SFR})$ joint posterior space of a galaxy with log(SFR)$\sim0$ and log$(M_\ast/M_\odot)\sim9$. We use this to generate 20,000 perturbations with means $\{\mu_x,~\mu_y\}$ = (0,0) and $\{\sigma_x, \sigma_y\}$ = (0.2,0.2), $\{dx_i,~dy_i\}$. We also generate 20,000 values for $\epsilon$ from a normal distribution, $\epsilon\sim$N($0,0.1$), to simulate intrinsic scatter along the linear relation. Adding this to our simulated set of data, we get $\{ \Tilde{x_i}, \Tilde{y_i} \}$ = $\{ x_i + dx_i, y_i + dy_i + \epsilon_i \}$. Finally, in order to ensure that we can accurately recover multiple values for intrinsic scatter along a linear relation, we scale the perturbations ($\{\epsilon_i\}$) to create two additional spaces with $\sigma=0.2$ and $\sigma=0.3$. We test our methodology on all three test cases, $\sigma=0.1,~0.2,~\mathrm{and}~0.3$, and the ``data'' are shown in Figure \ref{fig:app_fits}.

For each iteration of the MCMC simulation, we draw 2,000 samples from the set of 20,000 simulated ``galaxies'' to simulate sampling the SFR-$M_\ast$ posterior spaces of a real sample. Additionally, any simulated data with log(SFR)$\leq-0.4$ are excluded from the calculations. We do this to simulate an apparent selection bias against galaxies with low SFRs. We then compute the likelihood function for each walker, as described above. Figure \ref{fig:app_fits} shows the results of our fitting method, with a sample of 2000 simulated galaxies to demonstrate the scatter produced by the covariance matrix and intrinsic scatter for $\sigma=0.1,0.2~\&~0.3$ (left-to-right, respectively). The MCMC routine appears to trace the 2D distribution of simulated galaxies very well and we list the results in Table \ref{table:app_MCMC}. 

Figure \ref{fig:app_corner} shows the corner plots for each of the three test cases. As can be seen from both Figure \ref{fig:app_corner} and Table \ref{table:app_MCMC}, we underestimate $\sigma$ by $\sim30\%$ in all cases. Additionally, we are able to recover the $\alpha$ and $\beta$ in all cases. Although the intrinsic scatter along the relation is underestimated, the constant offset, $\sigma_{MCMC} = 0.7\sigma_{true}$, will not affect the results of our analysis. We could take the $\epsilon(z)$ values measured in our analysis and increase them by a factor of $1/0.7 \approx 1.4$ without affecting the results of this paper. However, we only note that that our results could be underestimated by this factor of $\sim1.4$ and do not apply the correction to any of the results presented in our analysis. In order to address this constant offset between the measured values and the ``True'' values, we also tested this method using $\alpha_{True}=0.8$, the value we measure for the SFMS in this analysis. We find that the constant offset remains, but decreases to $1/0.8 = 1.2$. We conclude that this constant offset is a result of an additional term that is introduced in $\{dx_i, dy_i + eta_i\}$ terms, causing an artificial tightening along the axis of the posterior spaces for galaxies. In the case of $\alpha_{True}=1$, this artificial tightening affect is greatest because the posterior space for the galaxy we use as our ``model'' posterior space is parallel to the simulated SFMS. Therefore, given the more randomly distributed orientation of the galaxy posterior spaces in our sample, this affect will be minimized through the random sampling of those posterior spaces. 

\begin{deluxetable*}{lllll}
\caption{Results of dual-likelihood MCMC testing}\label{table:app_MCMC}
\tablehead{\colhead{$\sigma_{True}=0.1$} & \colhead{} & \colhead{$\sigma_{True}=0.2$} & \colhead{} & \colhead{$\sigma_{True}=0.3$} }
\startdata
    $\hat{\alpha} =$\phs$1.01 \pm 0.02$ & & $\hat{\alpha} = 1.00\pm0.02$ & & $\hat{\alpha} =$\phs$1.01\pm0.02$  \\
    $\hat{\beta} = -0.01\pm0.02$ & & $\hat{\beta} = 0.00\pm0.03$ & & $\hat{\beta} = -0.01\pm0.03$ \\
    $\hat{\sigma} =$\phs$0.07\pm0.01$ & & $\hat{\sigma} = 0.14\pm0.01$ & & $\hat{\sigma} =$\phs$0.21\pm0.02$
\enddata
\end{deluxetable*}

\bibliography{citationsource}
\end{document}